\documentclass[aps,
               prd,
               twocolumn,
               superscriptaddress,
               amsfonts,
               amssymb,
               amsmath,
               preprintnumbers,
               nofootinbib,
               unsortedaddress,
               showpacs,
               tightenlines,
               floats,
               letterpaper,
               eqsecnum,
               ]{revtex4}

\usepackage{graphicx}
\usepackage{latexsym}
\usepackage{bm}
\usepackage[draft=false]{hyperref}

\def\F{{\mathcal F}}
\def\G{{\mathcal G}}

\def\O{{\mathcal O}}
\def\Q{{\mathcal Q}}
\def\R{{\mathcal R}}
\def\pbar{{\bar{p}}}
\def\nbar{{\bar{n}}}
\def\sun{{\odot}}
\def\Alfven{{Alfv\'{e}n}~} 

\begin{document}

\title{
Sub-GeV galactic cosmic-ray antiprotons from primordial black holes in the Randall-Sundrum braneworld
}

\author{Yuuiti Sendouda}\email[]{sendouda@utap.phys.s.u-tokyo.ac.jp}
\affiliation{Department of Physics, Graduate School of Science, The University of Tokyo, 7-3-1 Hongo, Bunkyo-ku, Tokyo 113-0033, Japan}
\author{Kazunori Kohri}
\affiliation{Department of Earth and Space Science, Graduate School of Science, Osaka University, 1-1 Machikaneyama, Toyonaka, Osaka 560-0043, Japan}
\author{Shigehiro Nagataki}
\affiliation{Yukawa Institute for Theoretical Physics, Kyoto University, Oiwake-cho, Kitashirakawa, Sakyo-ku, Kyoto 606-8502, Japan}
\author{Katsuhiko Sato}
\affiliation{Department of Physics, Graduate School of Science, The University of Tokyo, 7-3-1 Hongo, Bunkyo-ku, Tokyo 113-0033, Japan}
\affiliation{Research Center for the Early Universe, Graduate School of Science, The University of Tokyo, 7-3-1 Hongo, Bunkyo-ku, Tokyo 113-0033, Japan}

\date{
\today
}

\begin{abstract}
We investigate cosmic-ray antiprotons emitted from the galactic primordial black holes in the Randall-Sundrum type-2 braneworld.
The recent results of the BESS antiproton observation implies the existence of exotic primary sub-GeV antiprotons, one of whose most probable origin is PBHs in Our Galaxy.
We show that the magnitude of antiproton flux from PBHs in the RS braneworld is proportional to negative power of the AdS radius, and immediately find that a large extra-dimension can relax upper-limits on the abundance of the galactic PBHs.
If actually there are more PBHs than the known upper-limit obtained in the pure 4D case, they set a lower bound on the size of the extra dimension above at least $ 10^{20} $ times 4D Planck-length to avoid inconsistency.
On completion of the numerical studies, we show that these constraints on the AdS radius is comparable to those obtained from the diffuse photon background by some of the authors in the previous paper.
Moreover, in the low accretion-rate case, only antiprotons can constrain the braneworld.
We show that we will detect signatures of the braneworld as a difference between the flux of the antiprotons predicted in 4D and 5D by future observations in sub-GeV region with a few percent precision.
\end{abstract}



\preprint{UTAP-493}
\preprint{OU-TAP-233}
\preprint{YITP-04-43}
\preprint{RESCEU-30/04}

\maketitle

\section{Introduction}

The primordial black holes (PBHs), which are thought to be formed from density fluctuations during the radiation-dominated era of the early Universe \cite{Zel'dovichNovikov1966,Hawking1971,CarrHawking1974,Carr1975,Novikovetal1979,Bringmannetal2001,Blaisetal2003} (for other mechanisms, see also \cite{Rubinetal20002001}), have long been a useful tool to investigate the early Universe itself owing to their radiative nature discovered by Hawking \cite{Hawking19741975,Page1976,PageHawking1976,Carr1976,MiyamaSato1977,MacGibbon1987,MacGibbonCarr1991,KohriYokoyama1999,Barrauetal2003}.
A cosmological model under investigation in this paper is the so-called RS2 braneworld suggested by Randall and Sundrum.
In their scenario, our Universe is a 4D Lorentz metric hypersurface called {\it brane}, embedded in a $ Z_2 $ symmetric 5D anti--de Sitter (AdS) spacetime called {\it bulk} \cite{RS1999}.
RS2 assumes that the number of branes is only one and its tension is positive.
The bulk has a typical length scale that originates from AdS curvature radius, $ l $, which draws a short distance boundary on the brane; 4D Newton law is recovered as far as we consider a sufficiently larger scale than $ l $, while deviation from ordinary gravity appears below it \cite{RS1999,SMS2000}.
The present table-top upper bound on the bulk AdS radius $ l $ was obtained by measuring short-distance gravitational force as $ l \lesssim 0.2~\mathrm{mm} $ \cite{Longetal2003,Hoyleetal2001}.
From such a point of view, PBHs again retain their status.
Actually, in the framework of the braneworld, it has been revealed that the nature of black holes with small masses highly differs from the ordinary 4D one.
In particular small black holes with Schwarzschild radius $ r_\mathrm{S} \ll l $ behave as essentially 5D objects and their radii are significantly stretched compared with 4D BHs \cite{Emparanetal2000a,GiddingsThomas2002}.
It means that the temperature of Hawking radiation drops down.
We naturally expect that even such small black holes were produced by gravitational collapse in the early universe \cite{GCL2002a,Kawasaki2003} (there have been discussions on the black hole formation in respect of the particle scattering processes even in the early Universe \cite{GiddingsThomas2002,AhnCavaglia2003}).
Thus we have to revisit the particle production by primordial black holes taking into account how non-4D character modifies it.
In addition, the RS2 scenario implies a nonstandard high-energy regime before the standard radiation-dominated era \cite{Clineetal1999,Csakietal1999,Binetruyetal2000a,Binetruyetal2000b} (see, e.g., \cite{Langlois2003} and references therein).
It was recently argued by Guedens, Clancy, and Liddle and Majumdar \cite{GCL2002a,GCL2002b,Majumdar2003} that through the brane high-energy phase accretion of surrounding radiation onto PBHs is so efficient that they grow in their masses on the contrary to the standard scenario.
Actually it was already revealed that the high-energy diffuse photon background is highly modified in the braneworld \cite{CGL2003,SendoudaNagatakiSato2003}, and it can in turn constrains braneworld according to the amount of PBHs \cite{SendoudaNagatakiSato2003}.
Even in other various contexts, braneworld PBHs are gradually extending their status \cite{MajumdarMehtaLuck2003,MajumdarMukherjee2004}.

We extend above-mentioned strategy in this paper by incorporating cosmic-ray antiproton observations.
Because PBHs are thought to be emitting both particles and antiparticles without any distinction, we can search for dark matter PBHs with the aid of antimatter detectors.
In fact, one of the most precisely measured is antiproton among various antimatter species in cosmic rays.
High-precision measurements of antiprotons around GeV have been successively performed since mid-1990s by several collaborations such as Balloon-borne Experiment with a Superconducting Spectrometer (BESS) \cite{BESS1993,BESS1995,BESS1997,BESS1998}, Isotope Matter Antimatter eXperiment (IMAX) \cite{Mitchelletal1996}, Cosmic AntiParticle Ring Imaging Cerenkov Experiment (CAPRICE) \cite{CAPRICE1994}, and so on.
There are even observations in high energy regions over $ 3~\text{GeV} $ \cite{MASS2,CAPRICE1998}.
It is known that antiprotons can set nearly the same constraint on the PBH abundance as the diffuse gamma-ray in the 4D case \cite{MakiMitsuiOrito1996,MitsuiMakiOrito1996,Barrauetal2002}.
Therefore in this paper, we examine antiproton emission from braneworld PBHs and see the spectral shape of the antiproton flux.
Combined with the observational data, they give quantitative constraints on the amount of PBHs and, in the context of the braneworld, in turn set bounds on the AdS curvature radius $ l $.

For propagation of cosmic-ray antiprotons in the Galaxy, we take a recent version of the diffusion model \cite{Ginzburgetal1980,Webberetal1992,Maurinetal2001,Donatoetal2001,Maurinetal2002,BarrowMaartens2002,TailletMaurin2003,MaurinTaillet2003}.
Unfortunately, the effect of the solar wind is crucial for the sub-GeV antiproton flux.
Although such a effect, called solar modulation, has been under successive measurement, there remains a difficulty in incorporating it into the model.
Hence we wish to use data as little affected by the solar activity as possible, therefore choose BESS \cite{BESS1995,BESS1997,BESS1998} and CAPRICE \cite{CAPRICE1994,CAPRICE1998}, which were taken during the recent solar minimum.

The contents of this paper are as follows.
First we briefly review in Sec.~\ref{sec:review} the cosmology and behavior of evaporating black holes in the RS2 braneworld.
In Sec.~\ref{sec:formulation}, formulations to describe PBHs in the brane Universe are shown.
We analyze the dependence of the antiproton flux on the braneworld parameters in Sec.~\ref{sec:pbar}.
In this section, we also show numerical results for a wide range of braneworld parameters to give quantitative constraints on the PBH amount and themselves.
Discussions and conclusions are summarized in Sec.~\ref{sec:conclusion}.

\section{\label{sec:review}
Cosmology and black holes in the braneworld
}

\subsection{Braneworld cosmology}

Hereafter we basically use 4D natural units $ \hbar = c = k_\mathrm{B} = 1 $ throughout this paper.
Newton's gravitational constant is also taken to be unity and we define 4D Planck mass as $ M_4 \equiv G^{-1/2} = 1 $.
In these units, other 4D Planck-scale quantities, defined as $ l_4 $, $ t_4 $, and $ T_4 $, are also all unity, whereas they will usually appear in expressions to clarify their differences from higher dimensional fundamental scales.
A subscript ``5'' expresses five-dimensional quantity, such as 5D Planck mass $ M_5 $, and so on.

We start with the five-dimensional Einstein equation
\begin{equation}
{}^{(5)}R_{\mu\nu} - \frac{1}{2}{}^{(5)}R g_{\mu\nu}
 = \frac{8 \pi}{M_5^3} T_{\mu\nu},
\label{eq:brane.Einstein}
\end{equation}
where the energy-momentum tensor is expressed in the Gaussian normal coordinates as \cite{SMS2000}
\begin{equation}
T_{\mu\nu} = -\Lambda_5 g_{\mu\nu} + S_{\mu\nu} \delta(y), \quad
S_{\mu\nu} = \lambda q_{\mu\nu} + \tau_{\mu\nu},
\end{equation}
with the metric on the brane $ q_{\mu\nu} $.
Cosmological expansion on the brane is governed by the reduced 4D Einstein equation \cite{SMS2000},
\begin{equation}
{}^{(4)}G_{\mu\nu}
 = \Lambda_4 q_{\mu\nu}
 + \frac{8\pi}{M_4^2} \tau_{\mu\nu}
 + \frac{64\pi^2}{M_5^6} \pi_{\mu\nu}
 - E_{\mu\nu}
\end{equation}
with 
\begin{equation}
\Lambda_4
 = \frac{4 \pi}{M_5^3}
   \left(\frac{4 \pi}{3} \frac{\lambda^2}{M_5^3} - |\Lambda_5|\right), \quad
M_4^2
 = \frac{3}{4 \pi} \frac{M_5^6}{\lambda},
\end{equation}
and
\begin{equation}
\pi_{\mu\nu}
 = - \frac{1}{4} \tau_{\mu\alpha} \tau_\nu^\alpha
 + \frac{1}{12} \tau \tau_{\mu\nu}
 + \frac{1}{8} q_{\mu\nu} \tau_{\alpha\beta} \tau^{\alpha\beta}
 - \frac{1}{24} q_{\mu\nu} \tau^2.
\end{equation}
The standard perfect-fluid energy-momentum tensor on the brane $ \tau_{\mu\nu} $ leads to the effective 4D Friedmann equation.
In this framework, the expansion-law is modified from the pure 4D case due to the existence of the typical energy scale on the brane, the brane tension $ \lambda $ \cite{SMS2000,Clineetal1999,Csakietal1999,Binetruyetal2000a,Binetruyetal2000b}.
The result is
\begin{equation}
H^2
 = \frac{8\pi}{3 M_4^2}
   \left[\rho \left(1+\frac{\rho}{2\lambda}\right)+\rho_\mathrm{KK}\right]
 + \frac{\Lambda_4}{3}
 - \frac{K}{a^2}
\label{eq:brane.Friedmann.formal},
\end{equation}
where $ a $ is the scale factor, $ H $ is the Hubble constant, $ \rho $ and $ \rho_\mathrm{KK} $ are the energy density of ordinary matter field and of the dark radiation, respectively, and $ K $ is the intrinsic spatial curvature of the brane.
The energy conservation law is the same form as the 4D one:
\begin{equation}
\dot{\rho} + 3 H (\rho + p) = 0.
\end{equation}
For complete derivation of the effective Friedmann equations, see, e.g., \cite{Langlois2003} and references therein.

As for the dark radiation term, it was argued that this quantity must be sufficiently smaller than ordinary radiation at the time of nucleosynthesis \cite{Binetruyetal2000b,BarrowMaartens2002}.
Thus we can simply drop this term.
Additionally, based on WMAP observation \cite{Bennettetal2003}, we can set $ K = 0 $ and require $ \Lambda_4 $ to be extremely small.
Using the definition of the AdS curvature radius $ l \equiv \sqrt{3 M_5^3/|4 \pi \Lambda_5|} $, we obtain the relation between the geometrical scale and the brane tension:
\begin{equation}
l = \frac{3}{4\pi} \frac{M_5^3}{\lambda}.
\end{equation}
Therefore a following simple relation between the five-dimensional and four-dimensional Planck mass holds:
\begin{equation}
M_4^2 = l M_5^3.
\label{eq:brane.5dscale}
\end{equation}

As a result, the Friedmann Eq.~(\ref{eq:brane.Friedmann.formal}) becomes
\begin{equation}
H^2
 = \frac{8\pi}{3 M_4^2} \rho \left(1+\frac{\rho}{2\lambda}\right)
 + \frac{\Lambda_4}{3}.
\label{eq:brane.Friedmann}
\end{equation}
This expression differs from the ordinary 4D one with respect to the existence of a {\em $ \rho $-square} term.
This term is only relevant when the radiation energy is sufficiently denser than the brane tension, namely in the earliest era of the Universe.
Then we immediately understand that the cosmological evolution in the early Universe is split into two phases;
the earlier one is the nonstandard high-energy regime in which evolution of the scale factor $ a(t) $ obeys $ a(t) \propto t^{1/4} $, and the latter is the standard radiation-dominated Universe, with $ a(t) \propto t^{1/2} $.
The former is often called the {\em $ \rho $-square} phase.

In order to obtain whole profile of the scale factor, WMAP data, such as $ t_0=13.7~\text{Gyr} $, $ h=0.71 $, $ \Omega_{\mathrm{m},0}h^2=0.135 $, $ \Omega_{\Lambda,0}=0.73 $, and $ z_\mathrm{eq}=3233 $, are used.
The time of matter-radiation equality $ t_\mathrm{eq} $ is calculated to be $ 72.6~\mathrm{kyr} $ \cite{SendoudaNagatakiSato2003}.
There is the analytic solution through the whole radiation-dominated era:
\begin{equation}
a(t)
 = a_\mathrm{eq} \frac{t^{1/4}(t+t_\mathrm{c})^{1/4}}{t_\mathrm{eq}^{1/2}},
\end{equation}
where we denoted the time when transition occurs as $ t_\mathrm{c} \equiv l/2 $.
The asymptotic forms of the profile is determined as
\begin{subequations}
\begin{alignat}{4}
a(t)
 & = a_\mathrm{eq} \frac{t^{1/4} t_\mathrm{c}^{1/4}}{t_\mathrm{eq}^{1/2}}
 & \quad \text{for}
 & \quad t \ll t_\mathrm{c} \\
 & = a_\mathrm{eq} \frac{t^{1/2}}{t_\mathrm{eq}^{1/2}}
 & \quad \text{for}
 & \quad t_\mathrm{c} \ll t \leq t_\mathrm{eq}.
\end{alignat}
\end{subequations}
In terms of the Hubble constant
\begin{equation}
H(t)
 = \frac{2t+t_\mathrm{c}}{4t(t+t_\mathrm{c})},
\end{equation}
other cosmological quantities, energy density $ \rho(t) $, Hubble radius $ R_\mathrm{H}(t) $, and horizon mass $ M_\mathrm{H}(t) $, are expressed as
\begin{align}
\rho(t)
 & = \frac{3 M_4^2}{16\pi t_\mathrm{c}^2}
     \left[\sqrt{4 H(t)^2 t_\mathrm{c}^2 + 1} - 1\right], \\
R_\mathrm{H}(t)
 & = \frac{1}{H(t)}, \\
M_\mathrm{H}(t)
 & = \frac{M_4^2}{4 H(t)^3 t_\mathrm{c}^2}
     \left[\sqrt{4 H(t)^2 t_\mathrm{c}^2 + 1} - 1\right].
\end{align}
Hereafter we denote the horizon mass at the transition as $ M_\mathrm{c} \equiv M_\mathrm{H}(t_\mathrm{c}) = (16/27)l \sim l $, and call it {\em transition mass}.
We explicitly write down the asymptotic expressions for cosmological quantities in the early Universe for later use:
\begin{align}
 & \rho(t) = \frac{3 M_4^2}{32\pi t_\mathrm{c} t},
   \quad R_\mathrm{H}(t) = 4 t,
   \quad M_\mathrm{H}(t) = 8 M_4^2 \frac{t^2}{t_\mathrm{c}} \nonumber \\
 & \quad \text{for} \quad t \ll t_\mathrm{c}, \nonumber \\
 & \rho(t) = \frac{3 M_4^2}{32\pi t^2},
   \quad R_\mathrm{H}(t) = 2 t,
   \quad M_\mathrm{H}(t) = M_4^2 t \nonumber \\
 & \quad \text{for} \quad t \gg t_\mathrm{c}.
\label{eq:brane.cosmology}
\end{align}

\subsection{Five-dimensional black holes}

In the braneworld scenario, a massive object localized on the brane within typical length scale much shorter than $ \ll l $ should see the background space-time as an effectively flat 5D Minkowski space \cite{Wiseman2002,Kudohetal2003} (for possible deviations and other exotic features, see \cite{CasadioMazzacurati2003,CasadioHarms2002,Dadhichetal2000,DadhichiGhosh2001,BruniGermaniMaartens2001,GovenderDadhich2002}).
Therefore a sufficiently small black hole is no longer ordinary 4D one, but is approximately described as 5D solution\footnote{
The emergence of this topological change is not specific for the Randall and Sundrum's braneworld scenario, but a generic feature for higher-dimensional theories that have some {\it compactification} length scales in a broad sense.
We will briefly discuss this issue in the last of this paper.
See Sec.~\ref{sec:conclusion}.
}.
If that black hole has no charge and is not rotating, the solution must be 5D Schwarzschild.
In terms of asymptotically flat spherical coordinates, the 5D Schwarzschild metric is written down \cite{MyersPerry1986} and its 4D projection is
\begin{equation}
\mathrm{d}s^2
 = -C(r) \mathrm{d}t^2
 + \frac{1}{C(r)} \mathrm{d}r^2
 + r^2 \mathrm{d}\Omega_2^2
\label{eq:brane.metric}.
\end{equation}
with
\begin{equation}
C(r) = 1 - \left(\frac{r_\mathrm{S}}{r}\right)^2,
\end{equation}
where $ r_\mathrm{S} $ is the 5D Schwarzschild radius and $ \mathrm{d}\Omega_2 $ is the line element on unit $ 2 $-sphere.
In our framework the Schwarzschild radius in the five-dimensional spacetime is related to the black hole mass $ M $ and five-dimensional fundamental mass scale $ M_5 $ as \cite{MyersPerry1986}
\begin{align}
r_\mathrm{S}
 & = \left(\frac{8}{3\pi}\frac{M}{M_5^3}\right)^{1/2} \nonumber \\
 & = \sqrt{\frac{8}{3\pi}}
     \left(\frac{l}{l_4}\right)^{1/2}
     \left(\frac{M}{M_4}\right)^{1/2} l_4,
\label{eq:brane.radius}
\end{align}
while in the ordinary 4D case $ r_\mathrm{S} = 2 (M/M_4) l_4 $.
On the other hand, the Hawking temperature of a Schwarzschild black hole is determined by space-time periodicity in the direction of imaginary time.
For a higher-dimensional Schwarzschild black hole, the temperature is given in terms of its radius as $ T_\mathrm{H} = (D-3)/4\pi r_\mathrm{S} $ \cite{MyersPerry1986,GiddingsThomas2002}.
Thus the five-dimensional Hawking temperature is
\begin{equation}
T_\mathrm{H}
 = \frac{1}{2\pi r_\mathrm{S}}
 = \sqrt{\frac{3}{32 \pi}}
   \left(\frac{l}{l_4}\right)^{-1/2}
   \left(\frac{M}{M_4}\right)^{-1/2} T_4,
\label{eq:brane.temp}
\end{equation}
while in the 4D case $ T_\mathrm{H} = 1/4\pi r_\mathrm{S} = (1/8\pi) (M/M_4)^{-1} T_4 $.

\section{\label{sec:formulation}
Formulations
}

\subsection{Formation}

It has been argued that in the early Universe primordial black holes are formed due to gravitational collapse caused by density perturbation \cite{Zel'dovichNovikov1966,Hawking1971,CarrHawking1974,Carr1975,Novikovetal1979}.
We here simplify the situation;
when a superhorizon-scale density perturbation with adequate amplitude enters the Hubble horizon at $ t = t_\mathrm{i} $, it instantaneously collapses into a black hole with some fraction of the horizon mass.
Hence the initial mass of a PBH denoted $ M_\mathrm{i} $ is given by
\begin{equation}
M_\mathrm{i} = f M_\mathrm{H}(t_\mathrm{i}),
\label{eq:pbh.initialmass}
\end{equation}
where the $ \O(1) $ parameter $ f $ expresses above mentioned fraction \cite{GCL2002a,NiemeyerJedamzik1999}.
Utilizing the explicit forms of horizon mass in asymptotic regions, i.e., Eq.~(\ref{eq:brane.cosmology}), we can write down the relations between the time of formation and initial mass for two types of PBH as
\begin{subequations}
\begin{alignat}{4}
M_\mathrm{i}
 & = 16 f l^{-1} t_\mathrm{i}^2
 & \quad \text{for}
 & \quad t_\mathrm{i} \ll t_\mathrm{c} \\
 & = f t_\mathrm{i}
 & \quad \text{for}
 & \quad t_\mathrm{i} \gg t_\mathrm{c}.
\end{alignat}
\end{subequations}
Note that $ M_\mathrm{i} < f M_\mathrm{c} $ corresponds with 5D BHs while $ M_\mathrm{i} > f M_\mathrm{c} $ with ordinary 4D ones.

\subsection{Accretion}

Matter field localized on the brane sees a black hole as an approximately disk-shaped object with the effective Schwarzschild radius \cite{Emparanetal2000b}
\begin{equation}
r_{\mathrm{S},\mathrm{eff}}
 = \left(\frac{D-1}{2}\right)^{1/(D-3)} \left(\frac{D-1}{D-3}\right)^{1/2}
   r_\mathrm{S}.
\label{eq:pbh.effectiveradius}
\end{equation}
For $ D = 5 $ and $ D = 4 $, $ r_{\mathrm{S},\mathrm{eff}} = 2 r_{\mathrm{S}} $ and $ r_{\mathrm{S},\mathrm{eff}} = (3\sqrt{3}/2) r_\mathrm{S} $, respectively.

Consider an infinitesimal time interval $ [t,t+\mathrm{d}t] $.
Then the gain of black hole mass due to absorbing surrounding radiation within $ \mathrm{d}t $ is described as \cite{GCL2002b,Majumdar2003}
\begin{equation}
\left(\frac{\mathrm{d}M}{\mathrm{d}t}\right)_\mathrm{accr}
 = F \pi r_{\mathrm{S},\mathrm{eff}}^2 \rho(t),
\end{equation}
where $ \rho(t) $ is energy density of radiation.
Here we introduced a parameter $ F $, which determines the {\em accretion efficiency};
if the geometrical optics approximation holds well, namely an incidental relativistic particle can be regarded as a collisionless point particle and its spin can be ignored, $ F $ should be almost unity, otherwise smaller.
Due to the lack of knowledge of this issue, we are forced to treat $ F $ as a free parameter throughout discussions.

A significant point is here.
In the 4D case, the above differential equation does not have a solution which describes increasing mass.
However, in the brane high-energy era, it does give a growing solution;
{\em accretion} occurs in the braneworld scenario.
Substituting relevant expressions for the brane high-energy phase into the above differential equation, the resultant form becomes
\begin{equation}
\left(\frac{\mathrm{d}M}{\mathrm{d}t}\right)_\mathrm{accr}
 = \frac{2F}{\pi} \frac{M}{t} \quad \text{for} \quad t \lesssim t_\mathrm{c}.
\end{equation}
Integrating it gives the mass evolution formula \cite{GCL2002b,Majumdar2003}
\begin{equation}
M(t)
 = \left(\frac{t}{t_\mathrm{i}}\right)^{2 F/\pi} M_\mathrm{i} \quad \text{for} \quad t \lesssim t_\mathrm{c}.
\end{equation}
We define the {\em primordial} mass $ M_\mathrm{p} $ of 5D PBHs at the end of accretion as $ M_\mathrm{p} \equiv M(t_\mathrm{c}) $.
Readers should not confuse $ M_\mathrm{c} $ with $ M_\mathrm{p} $.
For 4D PBHs, $ M_\mathrm{p} \equiv M_\mathrm{i} $.

\subsection{\label{sec:pbh.evap}
Evaporation
}

The differential particle emission rate from a $ D $-dimensional static uncharged black hole with mass $ M $ is given by Hawking's formula:
\begin{equation}
\mathrm{d}\frac{\mathrm{d}N_j}{\mathrm{d}t}
 = g_j \frac{\sigma_j(M,E)}{\exp(E/T_\mathrm{H}) \pm 1}
   \frac{\mathrm{d}k^{D-1}}{(2 \pi)^{D-1}},
\label{eq:pbh.Hawking}
\end{equation}
where subscript $ j $ indicates particle species, $ g_j $ the number of internal degrees of freedom of the particle, $ E = \sqrt{k^2 + m_j^2} $ the energy including rest mass, $ \sigma_j $ the absorption cross section, the sign in the denominator expresses which statistics the particle obeys, and $ T_\mathrm{H} $ is the Hawking temperature.

The absorption cross section $ \sigma_j $ is an essential part of the famous quantity named {\em greybody} factor.
It determines how far the radiation spectrum deviates from that of blackbody.
Although black holes under our considerations are higher-dimensional, ordinary particles will be emitted by them into the four-dimensional brane, and once the Hawking temperature of a black hole is fixed, the emission processes are almost unchanged from the 4D case.
Therefore we can adopt the known energy and spin-dependence of absorption cross section $ \sigma_j $ obtained in the 4D case \cite{Page1976,MacGibbonWebber1990}.
It moderately oscillates in the lower energy region, but eventually approaches the geometrical cross section $ \pi r_\mathrm{S,eff}^2 $ for any particle species in the high-energy regime.

The life time of a black hole is calculated from the summation of all particles' emission rate.
Using the formula (\ref{eq:pbh.Hawking}) under relativistic limit, the total mass decreasing rate of a $ D $-dimensional Schwarzschild black hole becomes
\begin{align}
\left(\frac{\mathrm{d}M}{\mathrm{d}t}\right)_\mathrm{evap}
 & = -\sum_j \int\limits_0^\infty
   E \frac{\mathrm{d}^2N_j}{\mathrm{d}E \mathrm{d}t} \mathrm{d}E \nonumber \\
 & = -\sum_j \int\limits_0^\infty
   g_j \frac{\sigma_j(E)}{\exp(E/T_\mathrm{H}) \pm 1}
   \frac{\Omega_{D-2} E^{D-1}}{(2\pi)^{D-1}} \mathrm{d}E \nonumber \\
 & \approx -\sum_j \int\limits_0^\infty
   g_j \frac{A_{\mathrm{eff},D} \Omega_{D-3}/[(D-2) \Omega_{D-2}]}
            {\exp(E/T_\mathrm{H}) \pm 1} \nonumber \\
 & \quad \times \frac{\Omega_{D-2} E^{D-1}}{(2\pi)^{D-1}} \mathrm{d}E.
\end{align}
Here we assumed that the absorption cross sections $ \sigma_j $'s are independent of the kind of particle species and energy $ E $ under the limit.
Integration gives the Stefan-Boltzmann's law of evaporating black holes:
\begin{equation}
\left(\frac{\mathrm{d}M}{\mathrm{d}t}\right)_\mathrm{evap}
 \approx - g_D \sigma_D A_{\mathrm{eff},D} T_\mathrm{H}^D,
\end{equation}
where we introduced some quantities as follows.
$ g_D $ is an effective degree of freedom given by contributions from bosons and fermions:
\begin{equation}
g_D = g_{D,\mathrm{boson}} + \frac{2^{D-1}-1}{2^{D-1}} g_{D,\mathrm{fermion}}.
\end{equation}
$ \sigma_D $ is the $ D $-dimensional Stefan-Boltzmann constant\footnote{
This definition gives half of traditional four-dimensional Stefan-Boltzmann constant because $ g_{\text{photon}} = 2 $ was extracted.
}
\begin{equation}
\sigma_D = \frac{\Omega_{D-3}}{(D-2) (2 \pi)^{D-1}} \Gamma(D) \zeta(D),
\end{equation}
where $ \Gamma(D) $ and $ \zeta(D) $ are gamma and zeta functions, respectively.
$ A_{\mathrm{eff},D} $ is the $ (D-2) $-dimensional effective surface area
\begin{equation}
A_{\mathrm{eff},D} = \Omega_{D-2} r_{\mathrm{S},\mathrm{eff},D}^{D-2},
\end{equation}
where $ \Omega_{D-2} $ is the area of the $ (D-2) $-dimensional unit sphere and $ r_{\mathrm{S},\mathrm{eff},D} $ is the effective $ D $-dimensional Schwarzschild radius in Eq.~(\ref{eq:pbh.effectiveradius}) (with $ D $ explicitly shown).

In the braneworld, one should consider degrees of freedom on the brane and in the bulk separately because the former takes 3D phase space and 2D black hole surface area, while the latter takes 4D phase space and 3D surface area.
Thus the mass shedding formula takes the form below:
\begin{multline}
\left(\frac{\mathrm{d}M}{\mathrm{d}t}\right)_\mathrm{evap}
 = -g_\mathrm{matter} \tilde{A}_{\mathrm{eff},4} T_\mathrm{H}^4
   - g_\mathrm{graviton} A_{\mathrm{eff},5} T_\mathrm{H}^5 \\
\text{for} \quad r_\mathrm{S} \lesssim l.
\end{multline}
Note that for small five-dimensional black holes we must take $ \tilde{A}_{\mathrm{eff},4} = \pi r_{\mathrm{S},\mathrm{eff},5}^2 $.
$ g_\mathrm{matter} $ is the degree of freedom of ordinary matter, such as photons, neutrinos, electrons, quarks, and their antiparticles etc.
Even in the braneworld, it should take the same value as 4D one.
On the other hand, the graviton polarization degree of freedom reflects the dimensionality as $ g_\mathrm{graviton} = [D(D-3)]/2 $.
We take $ g_\mathrm{graviton} = 5 $ for 5D black holes.

What has to be carefully treated is the temperature-dependence of the effective degrees of freedom.
An increasing effective degree of freedom of a worming PBH will shorten the lifetime of itself.
This issue was investigated by MacGibbon and Webber \cite{MacGibbonWebber1990}, and MacGibbon \cite{MacGibbon1991}.
However, we can expect that the effects on our discussion will be limited because our main target is long-life PBHs and they have in general lower temperature than pion mass around $ 135~\text{MeV} $ or the so-called quark-hadron transition temperature.

Based on the above arguments, the mass-loss rate for a five-dimensional black hole is given as
\begin{multline}
\left(\frac{\mathrm{d}M}{\mathrm{d}t}\right)_\mathrm{evap}
 = -\frac{g_\mathrm{eff,5}}{2}
   \left(\frac{l}{l_4}\right)^{-1}
   \left(\frac{M(t)}{M_4}\right)^{-1} \frac{M_4}{t_4} \\
\quad \text{for} \quad M \lesssim f M_\mathrm{c}.
\label{eq:pbh.massloss}
\end{multline}
Here the value of the effective degree of freedom $ g_\mathrm{eff,5} $ is
\begin{align}
g_\mathrm{eff,5}
 & = \frac{1}{2.6} \times \left(\frac{1}{160}g_\mathrm{matter}
   + \frac{9\zeta(5)}{32 \pi^4}g_\mathrm{graviton}\right) \nonumber \\
 & \approx 0.023 \quad \text{for only massless}. \nonumber
\end{align}
The extra factor $ 1/2.6 $ in front of the algebraic factors is a consequence of numerical calculation carried out by Page in the 4D case \cite{Page1976}.
For later use, we also compute the 4D effective degrees of freedom, which is defined in the 4D evaporation equation as
\begin{equation}
\frac{\mathrm{d}M(t)}{\mathrm{d}t}
 = -\frac{g_\mathrm{eff,4}}{3} \left(\frac{M(t)}{M_4}\right)^{-2}
   \frac{M_4}{t_4}.
\end{equation}
4D standard counting for massless species gives $ g_\mathrm{eff,4} \approx 0.00075 $.

The solution for Eq.~(\ref{eq:pbh.massloss}) with an appropriate initial condition is easily obtained as
\begin{multline}
M(t)
 = \left[
    \left(\frac{M_\mathrm{p}}{M_4}\right)^2
    - g_\mathrm{eff,5}
      \left(\frac{l}{l_4}\right)^{-1}
      \left(\frac{t-t_\mathrm{c}}{t_4}\right)
 \right]^{1/2} M_4 \\
\quad \text{for} \quad M \lesssim f M_\mathrm{c}.
\label{eq:pbh.evap}
\end{multline}
From this formula we can estimate the lifetime of black hole $ t_\mathrm{life} $,
\begin{equation}
t_\mathrm{life}
 = g_\mathrm{eff,5}^{-1}
   \left(\frac{l}{l_4}\right)
   \left(\frac{M_\mathrm{p}}{M_4}\right)^2 t_4
 + t_\mathrm{c}.
\label{eq:pbh.lifetime}
\end{equation}
4D mass profile is
\begin{equation}
M(t)
 = \left[
    \left(\frac{M_\mathrm{i}}{M_4}\right)^3
    - g_\mathrm{eff,4}
      \left(\frac{t-t_\mathrm{i}}{t_4}\right)
 \right]^{1/3} M_4.
\end{equation}

As we will see later, all our attention is concentrated on the cases in which there are small five-dimensional PBHs with sufficiently longer lifetimes than the present age of the Universe.
For a typical 5D black hole with lifetime $ t_\mathrm{life} = t_0 = 13.7~\mathrm{Gyr} $, its primordial mass $ M_\mathrm{p}^* $ is calculated as
\begin{align}
M_\mathrm{p}^*
 = \sqrt{\frac{g_\mathrm{eff,5} t_0}{l}}
 = 3.0 \times 10^{9}
   \left(\frac{l}{10^{31} l_4}\right)^{-1/2} \quad \mathrm{g}.
\label{eq:M_c^*}
\end{align}
Thus the corresponding temperature is
\begin{equation}
T_\mathrm{H}^*
 = 57 \left(\frac{l}{10^{31} l_4}\right)^{-1/4} \quad \mathrm{keV}.
\label{eq:pbh.temp.now}
\end{equation}
For the consistency of the above discussion, there is another requirement that at least a black hole with $ M_\mathrm{p}^* $ is indeed five-dimensional, i.e., $ r_\mathrm{S}[M_\mathrm{p}^*] < l $.
It holds when $ l > 10^{20} l_4 $.
With $ l $ in the typical range $ 10^{20} l_4 $--$ 10^{31} l_4 $ of our interest, $ T_\mathrm{H}^* $ is
\begin{equation}
57~\mathrm{keV} < T_\mathrm{H}^* < 32~\mathrm{MeV},
\label{eq:Hawkingtemperature^*region}
\end{equation}
hence the consistency of low temperature approximation for those long-life 5D PBHs is confirmed.

\subsection{The PBH mass spectrum}

In order to evaluate the amount of PBHs in the Universe, we take a usual notation \cite{GreenLiddle1997}
\begin{equation}
\alpha_t(M_\mathrm{i})
 \equiv \frac{\rho_{\mathrm{PBH},M_\mathrm{i}}(t)}{\rho_\mathrm{rad}(t)},
\end{equation}
where $ \rho_{\mathrm{PBH},M_\mathrm{i}} $ is the mass density of PBHs with initial mass {\em on the order of} $ M_\mathrm{i} $ and $ \rho_\mathrm{rad} $ is the radiation energy density.
In this paper we only use its primordial value, i.e., 
\begin{equation}
\alpha_\mathrm{i}(M_\mathrm{i})
 \equiv \alpha_{t_\mathrm{i}}(M_\mathrm{i})
 = \frac{\rho_{\mathrm{PBH},M_\mathrm{i}}(t_\mathrm{i})}
        {\rho_\mathrm{rad}(t_\mathrm{i})}
 = \alpha_t(M_\mathrm{i}) \frac{a(t_\mathrm{i})}{a(t)},
\end{equation}
where $ t_\mathrm{i} $ is given for each $ M_\mathrm{i} $ from Eq.~(\ref{eq:pbh.initialmass}).
The upper limit on $ \alpha_\mathrm{i} $ will be written as $ \mathrm{LIM_i} $.

Using the above definitions, the comoving number density of PBHs with initial mass on the order of $ M_\mathrm{i} $ is given as
\begin{equation}
n_\mathrm{PBH}(M_\mathrm{i})
 = \alpha_\mathrm{i}(M_\mathrm{i})
   \frac{\rho_\mathrm{rad}(t_\mathrm{i})}{M_\mathrm{i}}
   a(t_\mathrm{i})^3 \theta(t_\mathrm{life} - t_0),
\label{eq:pbh.initialspectrum.def}
\end{equation}
where $ t_\mathrm{life} $ is the corresponding lifetime and $ \theta(t) $ is the theta function.
The relation between $ n_\mathrm{PBH}(M_\mathrm{i}) $ and the differential comoving mass spectrum $ \mathrm{d}n/\mathrm{d}M_\mathrm{i} $ can be written as \cite{SendoudaNagatakiSato2003}
\begin{equation}
n_\mathrm{PBH}(M_\mathrm{i})
 \sim M_\mathrm{i} \frac{\mathrm{d}n}{\mathrm{d}M_\mathrm{i}}.
\end{equation}
Some constraint on $ n_\mathrm{PBH}(M_\mathrm{i}) $ leads to an equivalent constraint on the mass spectrum $ \mathrm{d}n/\mathrm{d}M_\mathrm{i} $.
Thus throughout this paper, we use the notion of number density $ n_\mathrm{PBH} $ and mass spectrum $ \mathrm{d}n/\mathrm{d}M $ in parallel.
All the discussions can be freely converted by the relation
\begin{equation}
\frac{\mathrm{d}n}{\mathrm{d}M} = \frac{n_\mathrm{PBH}}{M}.
\end{equation}

The relevant mass spectrum for antiprotons, which is the issue of this paper, is that at the present moment of the Universe.
As is mentioned in Appendix~\ref{sec:pbh.presentspectrum}, there are four possible destinies for primordial black holes according to their primordial size.
In the large extra-dimension case $ l > 10^{20} l_4 $, the present spectrum is
\begin{align}
\frac{\mathrm{d}n}{\mathrm{d}M}
 & = \frac{3}{2^{17/4}\pi}
     \frac{a_\mathrm{eq}^3}
          {t_\mathrm{eq}^{3/2}}
     \alpha_\mathrm{i}[M_\mathrm{i}(M_\mathrm{p})]
     \left(1+\frac{8\F}{9}\right) 4^\F f^{1/8+\F} \nonumber \\
 &   \quad \times 
      l^{-3/8+\F} M
     \left[M^2+\frac{g_\mathrm{eff,5}t_0}{l}\right]^{-25/16-\F/2} \nonumber \\
 &   \qquad \qquad \text{for} \quad M_\mathrm{c}^* \gtrsim M \nonumber \\
 & = \frac{1}{16 \pi}
     \frac{a_\mathrm{eq}^3}
          {t_\mathrm{eq}^{3/2}}
     \alpha_\mathrm{i}[M_\mathrm{i}(M)] f^{1/2}
     \frac{g_\mathrm{eff,4}}{g_\mathrm{eff,5}} l M \nonumber \\
 &   \quad \times
     \left[\frac{g_\mathrm{eff,4}l}{g_\mathrm{eff,5}}
           \left(M^2-M_\mathrm{c}^2\right)
           + M_\mathrm{c}^3 + g_\mathrm{eff,4} t_0\right]^{-3/2} \nonumber \\
 &   \qquad \qquad \text{for} \quad M_\mathrm{c} \gtrsim M \gtrsim M_\mathrm{c}^*
     \nonumber \\
 & = \frac{3}{32 \pi}
     \frac{a_\mathrm{eq}^3}
          {t_\mathrm{eq}^{3/2}}
     \alpha_\mathrm{i}[M_\mathrm{i}(M)] f^{1/2}
     M^2 \left[M^3 + g_\mathrm{eff,4} t_0\right]^{-3/2} \nonumber \\
 &   \qquad \qquad \text{for} \quad M \gtrsim M_\mathrm{c},
\label{eq:pbh.presentspectrum}
\end{align}
where
\begin{equation}
\F \equiv \frac{9F}{8(\pi-F)}
\quad \text{and} \quad
M_\mathrm{c}^* \equiv \sqrt{M_\mathrm{c}^2-\frac{g_\mathrm{eff,5} t_0}{l}}.
\end{equation}
On the other hand in the small extra-dimension case $ l \lesssim 10^{20} l_4 $, those lightest PBHs with mass $ M \lesssim M_\mathrm{c}^* $ have already died out.

\section{\label{sec:pbar}
Antiprotons
}

In this section, we describe the generation and propagation of antiprotons in the Galaxy.
As our formulations are solely owed to previous works by some authors \cite{Barrauetal2002,Donatoetal2001,Maurinetal2001,Maurinetal2002,TailletMaurin2003,MaurinTaillet2003}, we here only briefly look back the results.
Some more details are in Appendix~\ref{sec:pbar.propagation}.
Almost all the differences from the ordinary 4D scenarios are Hawking temperature of the PBHs and their mass spectrum.

\subsection{Sources}

\subsubsection{\label{sec:pbar.pri}
Antiprotons from PBHs in the galactic halo
}

Now let us consider {\em primary} antiprotons emitted from PBHs distributed over the Milky-Way halo.
Hereafter we assume spherically symmetric density profile of PBHs as dark matter.
Based on computational studies on the evolution of dark matter density distribution, the present profile is parameterized in some ways.
One of its general expressions in the cylindrical coordinates ($ r = 0 $ is the galactic center and $ z = 0 $ corresponds to the disc) is
\begin{align}
\rho_\mathrm{PBH}(r,z)
 & = \rho_\mathrm{PBH}(R_\sun,0) \nonumber \\
 & \times \left(\frac{R_\sun}{\sqrt{r^2+z^2}}\right)^\gamma
    \left(\frac{R_\mathrm{core}^\alpha+R_\sun^\alpha}
               {R_\mathrm{core}^\alpha+(\sqrt{r^2+z^2})^\alpha}\right)^\epsilon,
\end{align}
where $ R_\mathrm{core} $ is the so-called core radius and $ R_\sun $ is the distance from the galactic center to the solar system \cite{NFW1996,Mooreetal1999}.
We here use $ \alpha = 2 $, $ \gamma = 0 $, $ \epsilon = 1 $, and $ R_\mathrm{core} = 3~\text{kpc} $.
Choice of parameters is expected not to significantly affect the results much \cite{Barrauetal2002}.
Which mass spectrum should be used is determined by how long it takes for the messages to propagate from evaporating PBHs up to the instrument on the earth.
For antiprotons, the time scale is sufficiently shorter than the cosmic age, hence we take the present mass spectrum $ \mathrm{d}n/\mathrm{d}M $ given in Eq.~(\ref{eq:pbh.presentspectrum}).
A mass spectrum has such a value that is averaged over the whole Universe, so now let us define the effective mass spectrum in the Galaxy $ \mathrm{d}\tilde{n}/\mathrm{d}M $ as
\begin{equation}
\frac{\mathrm{d}\tilde{n}}{\mathrm{d}M}(r,z,M)
 = \G \frac{\rho_\mathrm{PBH}(r,z)}{\rho_{\mathrm{PBH}(R_\sun,0)}}
   \frac{\mathrm{d}n}{\mathrm{d}M}(M),
\end{equation}
where $ \G $ represents the enhancement of the galactic matter density measured at the solar neighborhood relative to the extragalactic value.
It is said that $ \G \sim 10^5 $ \cite{GordonBurton1976}.

Next, we consider particle emission degrees of freedom from black holes of quarks, antiquarks, leptons, antileptons, and gauge bosons.
Among them, quarks, antiquarks, and gauge bosons are able to generate antiprotons via hadronization process in quark or gluon jets.
We make use of \textsc{pythia/jetset} \cite{Sjoestrand1994} for it.
Our procedure is not so different from the 4D situation;
the necessary ingredients that we must fix are only the Hawking temperature and Schwarzschild radius of a black hole.
Their only difference from the four-dimensional case is $ \O(1) $ numerical factor coming from the topological nature of the black hole space-time.
Hence in the calculation, we can trustfully use the 4D result of spin and energy-dependence of the cross section, which was explicitly shown by MacGibbon and Webber \cite{MacgibbonWebber1990}.
From here we denote the effective differential emission rate of antiprotons as $ \mathrm{d}^2\tilde{N}_\pbar/\mathrm{d}E\mathrm{d}t $ defined in the following manner:
\begin{multline}
\frac{\mathrm{d}^2\tilde{N}_\pbar}{\mathrm{d}E\mathrm{d}t}
 = \sum_j \int\limits_{E'=E}^\infty g_j
   \frac{\sigma_j(M, E')}{\exp(E'/T_\mathrm{H}) \pm 1} \\
 \times \left[
        \frac{\mathrm{d}g_{j\pbar}(E', E)}{\mathrm{d}E}
        + \frac{\mathrm{d}g_{j\nbar}(E', E)}{\mathrm{d}E}
        \right]
   \frac{E'^2}{2\pi^2} \mathrm{d}E',
\end{multline}
where subscript $ j = q,\bar{q},g,Z^0,W^\pm $ stands for quark or gauge boson species emitted from a black hole.
$ \mathrm{d}g_{j\bar{X}}/\mathrm{d}E $ is the differential fragmentation ratio into antihadrons for a particle $ j $.
We included antineutrons because they immediately decay into antiprotons.

\begin{figure}[ht]
\begin{center}
\scalebox{0.7}{
\includegraphics[width=12cm,clip]{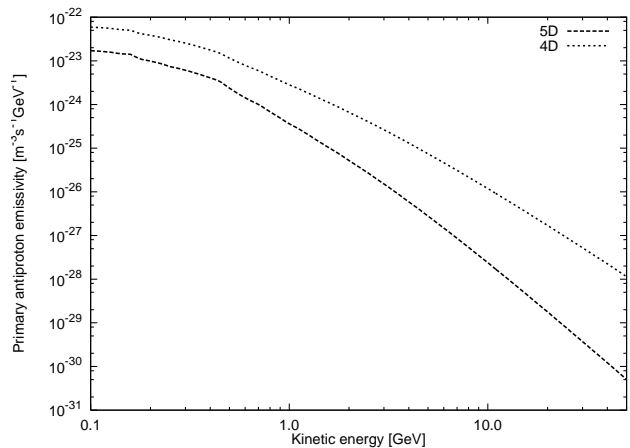}
}
\end{center}
\caption{\label{fig:pbar.q_pri}
Illustration of primary emissivity of antiprotons.
The difference between the two is mass spectrum of PBH.
Dotted line is 5D one in a typical case $ l = 10^{31} l_4 $, $ F = 1.0 $ and $ \G \alpha_\mathrm{i} = 10^{-22} $.
}
\end{figure}

From above considerations, we are now able to write down the differential antiproton emission rate from primordial black holes as
\begin{equation}
q^\mathrm{pri}(r,z,E)
 = \int \frac{\mathrm{d}^2\tilde{N}_\pbar}{\mathrm{d}E\mathrm{d}t}(M,E)
        \frac{\mathrm{d}\tilde{n}}{\mathrm{d}M}(r,z,M) \mathrm{d}M.
\label{eq:pbar.pri}
\end{equation}
Typical primary emissivity at the solar neighborhood is shown in Fig.~\ref{fig:pbar.q_pri}.

\subsubsection{\label{sec:pbar.sec}
Secondary antiprotons as the background
}

Next we consider {\it secondary} antiprotons.
The secondary antiprotons are generated in the particle reaction within the galactic disk between cosmic rays and the nuclei in the interstellar gas, such as $ \mathrm{N} + \mathrm{N}' \rightarrow \pbar + X $.
These process is thought to be well described by the known experimental data, so we treat secondaries as the background.
Actually this secondary component dominates the observed antiproton flux.
Because of the cosmic abundances, the most dominant sources are collisions between cosmic-ray protons and 90\% hydrogen plus 10\% helium in the interstellar gas \cite{Engelmannetal1990,Donatoetal2001}:
\begin{align}
p \mathrm{H}  & \rightarrow \pbar X, \nonumber \\
p \mathrm{He} & \rightarrow \pbar X. \nonumber
\end{align}
Proton cosmic ray has been measured by observations such as IMAX \cite{Mennetal2000}, BESS \cite{Sanukietal2000}, and Alpha Magnetic Spectrometer (AMS) \cite{Alcarazetal2000}.
The result of the observation of high-energy cosmic-ray protons at the solar neighborhood is well fitted by \cite{Donatoetal2001}
\begin{equation}
\Phi_p(R_\sun,E_p)
 = N \left(\frac{T_p}{1~\text{GeV}}\right)^{-\gamma}
   \quad \text{m$^{-2}$ sr$^{-1}$ s$^{-1}$ GeV$^{-1}$}
\end{equation}
with parameterizations $ N = 13249 $ and $ \gamma = 2.72 $, where $ T_p \equiv E_p - m_p $ is kinetic energy of protons.
Here we assume it is constant for $ r $ and use the above value.

As for the cross section, we make use of the most recent parameterization made by Duperray {\it et al.} \cite{Duperrayetal2003}, which gives better $ \chi^2 $ than the well-known result by Tan and Ng \cite{TanNg1982,TanNg1983}.
The differential cross section is given as
\begin{equation}
\frac{\mathrm{d}\sigma_{pA\rightarrow\pbar X}}{\mathrm{d}E_{\pbar}}
 = 2 \int\limits_{\theta_\mathrm{min}}^{\theta_\mathrm{max}}
   2 \pi p_\pbar \sin \theta_\pbar
   \left(E_\pbar \frac{\mathrm{d}^3\sigma}{\mathrm{d}p_\pbar^3}\right)
   \mathrm{d}\theta_\pbar,
\end{equation}
where $ p_{\pbar} = \sqrt{E_{\pbar}^2 - m_{\pbar}^2} $ and $ \theta_{\pbar} $ are momentum and azimuthal angle in the laboratory frame, respectively.
The limits of $ \theta_\pbar $ are determined by kinematics.
Again we multiplied the expression by 2 to incorporate antineutrons.

Finally, we obtain the production rate of the secondary antiprotons in the galactic disc as
\begin{multline}
q^\mathrm{sec}(r,0,E_\pbar)
 = \int\limits_{E_\mathrm{th}}^\infty \mathrm{d}E_p
   \left[4\pi\Phi_p(r, E_p)\right] \\
 \times  \sum_{A=\mathrm{H,He}} n_A
   \frac{\mathrm{d}\sigma_{pA \rightarrow \pbar X}}
        {\mathrm{d}E_\pbar}(E_p, E_\pbar),
\end{multline}
where the threshold of the interactions $ E_\mathrm{th} $ is $ 8 m_p $.
We use $ n_\mathrm{H} = 1~\text{cm}^{-3} $ and $ n_\mathrm{He} = 0.1~\text{cm}^{-3} $ for number the densities of the targets \cite{GordonBurton1976}.

\subsubsection{\label{sec:pbar.ter}
The tertiary source
}

There is another source of antiprotons, so-called {\it tertiary} component.
This is not a real production mechanism of antiproton but just energy exchanges between propagating antiprotons and ambient protons in the galactic disc.
The relevant interaction is the resonant excitation of the proton, which is non-annihilating but inelastic \cite{TanNg1983}.
The tertiary component is expressed as \cite{Donatoetal2001}
\begin{multline}
q^\mathrm{ter}(r,0,E_\pbar) \\
 = (n_\mathrm{H} + 4^{2/3}n_\mathrm{He})
   \int\limits_{E_\pbar}^\infty
   \frac{\sigma_{\pbar p}^\mathrm{non-ann}(T_\pbar')}
        {T_\pbar'}
   \left[4\pi \Phi_\pbar(r,0,E_\pbar')\right] \mathrm{d}E_\pbar' \\
 - (n_\mathrm{H} + 4^{2/3}n_\mathrm{He})
   \sigma_{\pbar p}^\mathrm{non-ann}(T_\pbar)
   \left[4\pi \Phi_\pbar(r,0,E_\pbar)\right].
\end{multline}
Due to the weakness of those interactions, the tertiary source is left ignored until numerical analyses.

\subsection{\label{sec:pbar.analysis}
Analyses on antiproton flux
}

It is obvious that $ \pbar $'s are only emitted from PBHs with typical temperature $ T_\mathrm{H} \gtrsim m_\pbar \sim 1~\text{GeV} $.
The mass corresponding with $ T_\mathrm{H} \sim 1~\text{GeV} $, $ M_\mathrm{GeV} $, is from Eq.~(\ref{eq:brane.temp})
\begin{alignat}{4}
M_\mathrm{GeV}
 & \sim 10^{14} \left(\frac{l}{10^{20}}\right)^{-1} \quad \text{g}
 & \quad \text{for}
 & \quad l \gtrsim 10^{20} \\
 & \sim 10^{14} \quad \text{g}
 & \quad \text{for}
 & \quad l \lesssim 10^{20}.
\end{alignat}
A condition $ M_\mathrm{GeV} < M_\mathrm{p}^* $ holds in any situation, which means that PBHs being such hot by nature, i.e., since their formation, cannot survive till now.
Thus contributions to the antiproton flux are all from evaporating region in the present mass spectrum, and we further find that the contribution to antiproton flux is dominated by only those PBHs with mass around $ M_\mathrm{GeV} $.
As a consequence, we understand that no braneworld signature in the antiproton flux is expected when $ l \lesssim 10^{20} $ because then the shape of the present mass spectrum around $ M_\mathrm{GeV} $ does not deviate at all from that in the 4D case.
On the other hand, when $ l \gtrsim 10^{20} $, $ M_\mathrm{GeV} $ PBHs are five dimensional and the effect of the braneworld will be strongly reflected in the $ \pbar $ flux.

\begin{figure}[ht]
\begin{center}
\scalebox{0.7}{
\includegraphics[width=12cm,clip]{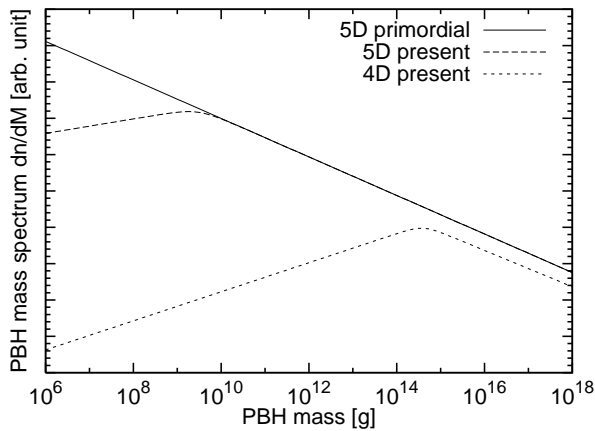}
}
\end{center}
\caption{\label{fig:pbar.ms}
Lines are 5D PBH mass spectrum soon after accretion (solid), 5D present one (dotted), and present one in the 4D case (dashed), respectively.
Braneworld parameters are taken to be $ l = 10^{31} l_4 $ and $ F = 1.0 $.
}
\end{figure}

Combining with another fact that the $ \pbar $ emission rate per a PBH is completely determined by its Hawking temperature only, one reaches the conclusion:
the spectral shape, such as peak location of the resultant antiproton flux, is never changed even if $ l $ or $ F $ varies, while the only possible modification is over-all fluctuation.
Under the low-mass limit $ M_\mathrm{GeV} \ll M_\mathrm{c}^* $, we can schematically write down the dependence of the observable $ \pbar $ flux at the solar neighborhood, $ \Phi_\pbar $, on the mass spectrum as
\begin{align}
\Phi_\pbar
 & \propto \int\limits^{M_\mathrm{GeV}} \mathrm{d}M
           \frac{\mathrm{d}^2\tilde{N}}{\mathrm{d}E \mathrm{d}t}
           \frac{\mathrm{d}\tilde{n}}{\mathrm{d}M} \nonumber \\
 & \sim \int\limits^{M_\mathrm{GeV}} \mathrm{d}M
     \frac{\mathrm{d}^2\tilde{N}}{\mathrm{d}E \mathrm{d}t}
     \frac{M}{M_\mathrm{GeV}}
     \left.\frac{\mathrm{d}\tilde{n}}{\mathrm{d}M}\right|_{M_\mathrm{GeV}} \nonumber \\
 & \propto M_\mathrm{GeV}
           \left.\frac{\mathrm{d}\tilde{n}}
                      {\mathrm{d}M}\right|_{M_\mathrm{GeV}}, 
\end{align}
with the mass spectrum from Eq.~(\ref{eq:pbh.presentspectrum3}):
\begin{multline}
\left.\frac{\mathrm{d}n}{\mathrm{d}M}\right|_{M_\mathrm{GeV}}
 \approx \frac{3}{2^{17/4}\pi}
         \frac{a_\mathrm{eq}^3}
              {t_\mathrm{eq}^{3/2}}
         \alpha_\mathrm{i}[M_\mathrm{i}(M_\mathrm{p})]
         \left(1+\frac{8\F}{9}\right) \\
 \times 4^\F f^{1/8+\F} l^{-3/8+\F} M_\mathrm{GeV}
        \left[\frac{g_\mathrm{eff,5}t_0}{l}\right]^{-25/16-\F/2} \\
 \propto \alpha_\mathrm{i} l^{3/16+3\F/2}
         \left[g_\mathrm{eff,5}t_0\right]^{-\F/2},
\end{multline}
where $ M_\mathrm{GeV} \propto l^{-1} $.
Finally, the parameter dependence of the flux is obtained as
\begin{equation}
\Phi_\pbar
 \propto \G \alpha_\mathrm{i} l^{-13/16+3\F/2}
           \left[g_\mathrm{eff,5}t_0\right]^{-\F/2}.
\label{eq:pbar.flux}
\end{equation}
Here we show the evolution of PBH mass spectrum in a typical case $ l = 10^{21} l_4 $ and $ F = 1.0 $ as in Fig.~\ref{fig:pbar.ms}.

The most important result is that in general the antiproton flux is a decreasing function of $ l $.
Recalling that $ \F = 9F/(8\pi-8F) $ takes a value in the range of $ 0 \leq \F \leq 0.53 $, it is understood that the index of $ l $, which is denoted as $ p $, is in fact always negative: $ p \in [-0.8125, -0.0175] $.
Hence we find that abundant PBHs set a {\em lower bound} on the bulk curvature radius $ l $.
In another word, the limits on $ \alpha_\mathrm{i} $ is proportional to $ l^{13/16-3\F/2} $, which is an increasing function of $ l $.
{\em Large $ l $ can relax the upper bound on $ \alpha_\mathrm{i} $}.

For completeness, we examine how the $ \F $ dependence is.
Taking a logarithm of Eq.~(\ref{eq:pbar.flux}), we obtain
\begin{align}
\log \Phi_\pbar
 \propto \log \left[\G \alpha_\mathrm{i}\right]
   - \left(\frac{13}{16}-\frac{3\F}{2}\right) \log l
   - \frac{\F}{2}
     \log \left[g_\mathrm{eff,5}t_0\right].
\end{align}
When the flux reaches some upper bound, which means that $ \alpha_\mathrm{i} $ also reaches its upper limit $ \mathrm{LIM_i} $, $ l $ has the minimum value $ l_\mathrm{min} $.
Then the above proportionality becomes
\begin{multline}
\log \left[\mathrm{U.L.}\right]
 = \log \left[\G \mathrm{LIM_i}\right] \\
 - \left(\frac{13}{16}-\frac{3\F}{2}\right) \log l_\mathrm{min}
 - \frac{\F}{2}
   \log \left[g_\mathrm{eff,5}t_0\right],
\end{multline}
where U.L. is an undetermined constant which corresponds with the upper limit flux.
In principle U.L. cannot be determined unless the propagation process is solved.
However, even without the knowledge of propagation, we can find more about the above relation.
What is important is the fact that the only difference for each case is the over-all fluctuation of the antiproton flux.
Therefore we can understand that U.L. is a {\it universal} value.
Keeping it in mind, we modify the relation as
\begin{multline}
\log l_\mathrm{min}
 = \frac{\log [g_\mathrm{eff,5} t_0]}{3} \\
 \times  \left[1+\frac{48\log[\G \mathrm{LIM_i}/\text{U.L.}]
                 /\log [g_\mathrm{eff,5} t_0]-13}
                {13-24\F}\right].
\label{eq:pbar.boundary}
\end{multline}
Although we cannot determine in general the sign of the numerator in the right hand side, we are particularly interested in the cases of large PBH abundance, which means that the numerator is positive.
For the same reason as the flux is a decreasing function of $ l $, the denominator is always positive for $ 0 \leq \F \leq 0.53 $.
Therefore we finally find that the lower bounds for $ l $ draw a family of hyperbolas expressed in Eq.~(\ref{eq:pbar.boundary}).

\subsection{\label{sec:pbar.numerical}
Numerical studies
}

Our final task is to give quantified conclusions by a numerical study on antiproton propagation in the Galaxy.
We adopt here a semi-analytical method which was first proposed by Webber, Lee, and Gupta \cite{Webberetal1992} and has been extensively utilized by other authors \cite{Barrauetal2002,Donatoetal2001,Maurinetal2001,Maurinetal2002,TailletMaurin2003,MaurinTaillet2003}.

\subsubsection{Diffusion}

This method is based on the diffusion model which was proposed by Ginzburg, Khazan, and Ptuskin \cite{Ginzburgetal1980}.

The full diffusion equation in the steady-state for the differential number density $ N(r,z,E) \equiv \mathrm{d}n_\pbar/\mathrm{d}E $ with the source $ \Q $ is \cite{Berezinskiietal1990}
\begin{multline}
0 = \frac{\partial N}{\partial t}
  = \vec{\nabla}
    \cdot [K(E) \vec{\nabla} N(r,z,E) - \vec{V}_\mathrm{c}(r,z) N(r,z,E)] \\
  + \frac{\vec{\nabla}\cdot\vec{V}_\mathrm{c}(r,z)}{3}
    \frac{\partial}{\partial E}
    \left[\frac{p^2}{E}N(r,z,E)\right]
  - \Gamma(E) N(r,z,E) \\
  + \Q(r,z,E)
  + \frac{\partial}{\partial E}
    \left\{-\left[b_\mathrm{reacc}(E)+b_\mathrm{coll}(E)\right] N(r,z,E)\right\} \\
  + \frac{\partial}{\partial E}
    \left[\beta^2 K_{pp}(E) \frac{\partial}{\partial E} N(r,z,E)\right].
\end{multline}
$ K $ and $ K_{pp} $ are the diffusion coefficient in the real space and energy space, respectively.
We assume power-law for the coefficient as $ K = K_0 \beta p^\delta $ with $ \delta = 0.6 $ \cite{Maurinetal2001,Donatoetal2001,Maurinetal2002,BarrowMaartens2002,TailletMaurin2003,MaurinTaillet2003}.
In terms of $ K $, we put $ K_{pp} \propto V_\mathrm{A}^2 p^2/K $, where $ V_\mathrm{A} $ is the \Alfven velocity.
$ \vec{V}_\mathrm{c} $ is the velocity of convective wind.
$ \Gamma $ expresses the escape probability from an energy bin via general inelastic interactions, while $ b_\mathrm{readd} $ and $ b_\mathrm{coll} $ correspond with energy losses by reacceleration and collision, respectively.

We put a cylindrically symmetric region, where diffusion occurs.
It is separated into two parts: the thin galactic disc, and the so-called diffusion halo sandwiching the disc.
The radius of the cylinder, $ R $, is set to $ 20~\text{kpc} $ and the solar system is located at $ R_\sun = 8~\mathrm{kpc} $ from the galactic center \cite{StanekGarnavich1998,Alves2000}.
The half-height of the disc, denoted as $ h $, is fixed at $ 100~\text{pc} $, while that of the diffusion halo, $ L $, is set to $ 6~\text{kpc} $.
Owing to the cylindrical symmetry, we express all the physical quantities that have radial dependence in terms of the Bessel functions of order zero, so that the diffusion equation is reduced to those for the Bessel components.
After appropriate evaluation of the terms, which is presented in Appendix~\ref{sec:pbar.propagation}, the final form of the diffusion equation for the antiproton number density becomes
\begin{multline}
0 = \left[
 K(E) \frac{\partial^2}{\partial z^2}
 - V_\mathrm{c} \frac{\partial}{\partial z}\right] N_i(z,E) \\
 - \left[K(E) \frac{\zeta_i^2}{R^2}
          + 2 h \delta(z) \Gamma_{\pbar p}(E)\right] N_i(z,E) \\
 + \Q_i(z,E)
 + \left\{\frac{\partial}{\partial E}
         \left[-b(E) + \beta^2 K_{pp}(E)
                        \frac{\partial}{\partial E}\right]
   \right\} N_i(z, E). \\
(i = 1,2,\ldots)
\end{multline}
$ b = b_\mathrm{reacc} + b_\mathrm{coll} + b_\mathrm{adiab} $, where $ b_\mathrm{adiab} $ corresponds with energy loss by adiabatic expansion.
This is the basic equation to be solved.

\subsubsection{\label{sec:pbar.solar}
Solar modulation
}

An obstacle for antiprotons approaching the earth is the solar wind.
The solar wind is a mixture of plasma and frozen-in magnetic field spouting from the sun.
As a result of propagation against the wind, a significant part of antiproton energy will be lost and the flux measured at the top of atmosphere (TOA) on the earth will be suppressed relative to the value at interstellar (IS) space.
This phenomenon is called solar modulation, which is synchronized with the solar activity with 11 years periodicity.
The most energetic period of the sun is called solar maximum, while the least energetic is solar minimum.

Although there are several ways to treat solar modulation, it has been known that they have their merits and demerits \cite{Asaoka2002}, so we here use a basic method, so-called force field approximation.
Basically, in every framework for the process, an electrical-potential-like quantity $ \phi $, which has a unit of GV. is intoduced to express the difficulty for cosmic-rays to flow against the solar wind.
Actually the force-field approximation method does.
In practice, $ \phi $ is estimated by the observations of cosmic-ray proton or neutron flux.
This parameter takes a value from $ 500~\mathrm{MV} $ at the solar minimum to $ 1~\mathrm{GV} $ at the solar maximum.
In the model of force-field approximation, the resultant energy at TOA is determined by that at IS and $ \phi $ as
\begin{equation}
\frac{E^\mathrm{TOA}}{A}
 = \frac{E^\mathrm{IS}}{A} - \frac{|Ze|\phi}{A},
\end{equation}
where $ A $ and $ Z $ are mass and the charge number of the cosmic-ray content;
for antiprotons they are both set to unity.
The ratio of the flux is written in terms of their momentum as
\begin{equation}
\frac{\Phi^\mathrm{TOA}}{\Phi^\mathrm{IS}}
 = \left(\frac{p^\mathrm{TOA}}{p^\mathrm{IS}}\right)^2.
\end{equation}

The solar modulation is so crucial for our discussion because it generally results in the energy loss ranging from a few hundred MeV at solar minimum to over GeV at solar maximum.
Hence most signatures of PBH evaporation, typically appearing in sub-GeV region, will be swept out at solar maximum.
Therefore we should take solar minimum data to be compared with the theoretical framework.
This is the reason why we will concentrate on BESS and CAPRICE data at the solar minimum \cite{BESS1995,BESS1997,BESS1998,CAPRICE1994,CAPRICE1998}.

\subsubsection{Semi-analytical method}

The diffusion equation can be regarded as a composite of the two qualitatively distinctive parts, i.e., diffusion in the real space and that in the energy space.
Basically, the energy part can be treated as a perturbation.

To integrate such a type of differential equation, we use a semi-analytical method \cite{Donatoetal2001}.
First we integrate the equation in the direction of $ z $, omitting both energy gain/loss terms and the tertiary source.
We use the annihilation cross section for $ \Gamma_{\pbar p} $ here (see Appendix~\ref{sec:pbar.propagation.interaction}).
Then that solution is substituted into resulting perturbative differential equation in the energy space together with the tertiary source term, and we solve it numerically\footnote{In this final process, we substitute the total cross section for inelastic interaction into $ \Gamma_{\pbar p} $.}.
We will iterate the energy diffusion till the solution converges.

\subsubsection{Choice of propagation parameters}

In order to clarify the issue, we shall fix values for the free parameters:
three diffusion parameters and one solar modulation parameter, i.e., $ K_0 $, $ V_\mathrm{A} $, $ V_\mathrm{c} $, and $ \phi $.
We understand that the parameters should be in principle determined independently of the braneworld.
Moreover, our purpose is solely focused on seeing how brane parameters affect the antiproton flux.
It means that models for propagation is here only required to be {\it reasonable}, thus we do not have to repeat fluctuating propagation parameters and searching the least $ \chi^2 $ for every braneworld parameter set.
It is sufficient for our purpose to fix reasonable values for the above parameters.

First we consider diffusion.
The three quantities should be in principle determined by observations, but unfortunately there has been no direct, precise observational result.
Thus the only trustful way to fix them is fitting cosmic-ray data to the theory \cite{Donatoetal2001,Barrauetal2002}.
In the procedure, basically we must try to exclude uncertainties originating from cosmic-ray sources otherwise they will introduce new errors for the parameters.
From such a point of view, antiproton is not a very appropriate cosmic-ray nucleus because we have been taking a standpoint such that their production rate is largely varied by both the amount of PBHs and braneworld parameters.
Therefore we borrow a result obtained by the analyses on the cosmic-ray carbon flux to boron (B/C) ratio \cite{Maurinetal2002,TailletMaurin2003,MaurinTaillet2003}.
Although the authors could not reach the goal to get the best-fit parameter set, they succeeded to show sufficiently reasonable parameter regions.
Our choice for the parameters, which is shown in Tab.~\ref{tab:pbar.param}, is consistent with their result.

The next problem is the solar modulation.
As was mentioned before, we use only solar minimum data within a few years.
Although each author having performed every year's observation has determined the best-fit $ \phi $ for the solar activity, their method to model the solar modulation is respectively different;
especially, some are not the same as our choice, the force-field approximation.
Hence we cannot directly use their values.
Here we take a conservative value $ 500~\mathrm{MV} $ for $ \phi $.

\begin{table}[ht]
\caption{\label{tab:pbar.param}
Propagation parameters sufficiently good for our purpose.
They are all consistent with previous results obtained for ordinary cosmic rays
\cite{Maurinetal2002,TailletMaurin2003,MaurinTaillet2003}.
}
\begin{center}
\begin{tabular}{ccccc}
\hline\hline
Quantity
 & $ K_0 $
 & $ V_\mathrm{A} $
 & $ V_\mathrm{c} $
 & $ \phi $ \\
\hline
Value
 & $ 0.012~\text{kpc$^2$ Myr$^{-1}$} $
 & $ 60~\text{km s}^{-1} $
 & $ 15~\text{km s}^{-1} $
 & $ 500~\text{MV} $ \\
\hline\hline
\end{tabular}
\end{center}
\end{table}

\subsubsection{\label{sec:pbar.region}
Allowed regions
}

\begin{figure}[ht]
\begin{center}
\scalebox{0.7}{
\includegraphics[width=12cm,clip]{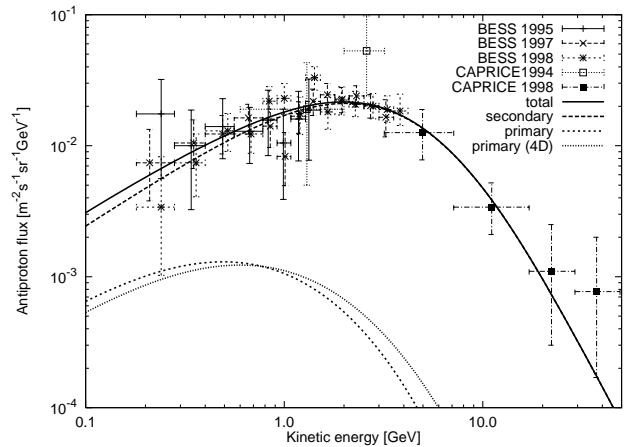}
}
\end{center}
\caption{\label{fig:pbar.total}
Total antiproton flux with a typical braneworld parameter set $ l = 10^{31}l_4 $ and $ F = 1.0 $.
Primary flux is the best-fit one ($ \G \alpha_\mathrm{i} = 1.1 \times 10^{-23} $).
4D best-fit primary ($ \G \alpha_\mathrm{i} = 1.8 \times 10^{-24} $) is also shown for comparison (4D total is not displayed).
}
\end{figure}

The final project is to obtain upper limits on the abundance of PBHs for each braneworld parameter set $ (l,F) $.
We fix the diffusion parameters and only let the PBH abundance $ \G \alpha_\mathrm{i} $ move.
First we show spectra with brane parameters $ l = 10^{31}l_4 $ and $ F = 1.0 $ in Fig.~\ref{fig:pbar.total}, which are corresponding to the best-fit $ \G \alpha_\mathrm{i} $.

From Fig.~\ref{fig:pbar.total}, we find that a few percent precision observation in sub-GeV region can distinguish the two primary flux, {\it i.e.}, we will be able to test the braneworld with future observations.

\begin{figure}[ht]
\begin{center}
\scalebox{0.7}{
\includegraphics[width=12cm,clip]{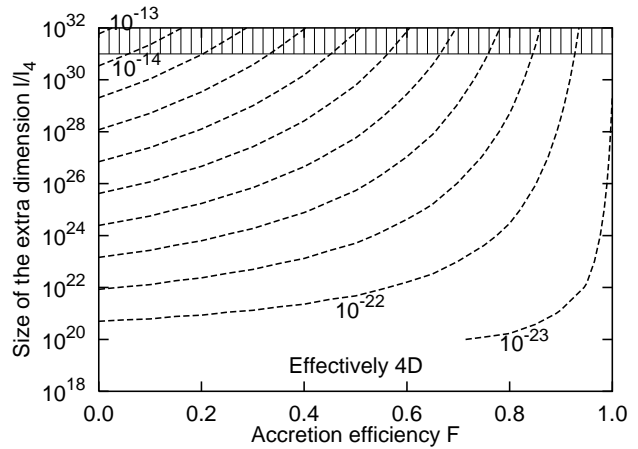}
}
\end{center}
\caption{\label{fig:pbar.region.0}
Boundaries of $ (l,F) $ in the braneworld-parameter plane for PBH abundance $ \G \alpha_\mathrm{i} = 10^{-13} \text{--} 10^{-23} $ (from top to bottom) using best-fit parameter.
They are all lower bounds of $ l $, i.e., left-top regions are allowed for each PBH abundance.
}
\end{figure}

\begin{figure}[ht]
\begin{center}
\scalebox{0.7}{
\includegraphics[width=12cm,clip]{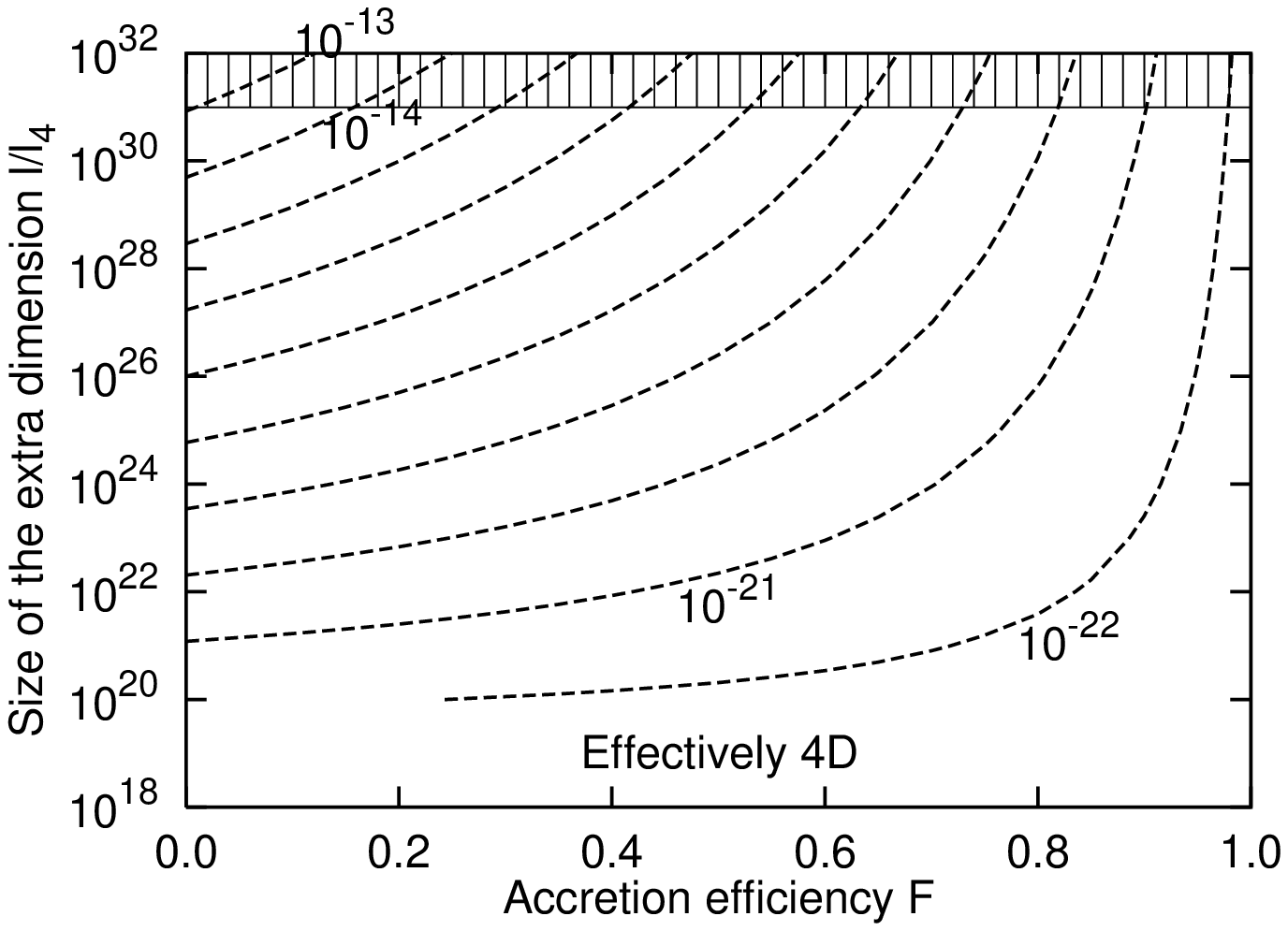}
}
\end{center}
\caption{\label{fig:pbar.region.1}
Boundaries of $ (l,F) $ for $ \G \alpha_\mathrm{i} = 10^{-13} \text{--} 10^{-22} $ (from top to bottom) using 50\% C.L. data.
}
\end{figure}

\begin{figure}[ht]
\begin{center}
\scalebox{0.7}{
\includegraphics[width=12cm,clip]{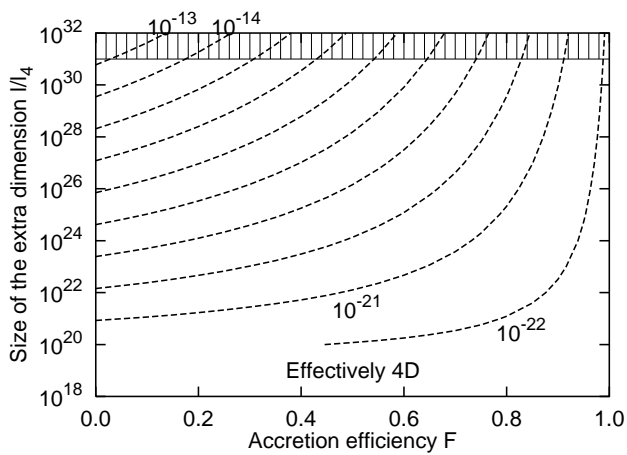}
}
\end{center}
\caption{\label{fig:pbar.region.2}
Boundaries of $ (l,F) $ for $ \G \alpha_\mathrm{i} = 10^{-13} \text{--} 10^{-22} $ (from top to bottom) using 90\% C.L. data.
}
\end{figure}

The final results of this section are allowed regions in the parameter plane $ (l,F) $ for best-fit (Fig.~\ref{fig:pbar.region.0}), 50\% C.L. (Fig.~\ref{fig:pbar.region.1}), and 90\% C.L. (Fig.~\ref{fig:pbar.region.2}).

At this stage, we can make a crude estimation of the cosmological PBH abundance.
We see in Figs.~\ref{fig:pbar.region.0}--\ref{fig:pbar.region.2} that non-trivial results are obtained for galactic PBH abundance ranging in the region $ 10^{-12} \gtrsim \G \alpha_\mathrm{i} \gtrsim 10^{-23} $.
Although $ \G $ is not determined yet, its order of magnitude is around $ 10^5 $, so the results are indicating that $ \mathrm{LIM_i} \sim 10^{-17} \text{--} 10^{-28} $ for $ l \gtrsim 10^{20} l_4 $.
Nothing happens in the case $ \alpha_\mathrm{i} \lesssim 10^{-29} $.
Comparing the constraint with that in the diffuse photon case \cite{SendoudaNagatakiSato2003}, we find that they are almost the same order constraints.

We here comment on the result obtained in the 4D setup.
The best-fit PBH abundance is $ \G \alpha_\mathrm{i} \approx 1.8 \times 10^{-24} $, which is consistent with the result obtained by Barrau {\it et al.} \cite{Barrauetal2002}.
When we assume $ \G \sim 10^5 $, $ \alpha_\mathrm{i} $ becomes around $ 10^{-29} $.
Although there are large uncertainties, it is about two orders of magnitude lower than the upper bound obtained from the diffuse gamma-ray background \cite{GreenLiddle1997,SendoudaNagatakiSato2003}.

\section{\label{sec:conclusion}
Discussions and conclusions
}

After the analyses, free parameters $ f $, $ F $, and $ \alpha_\mathrm{i} $ are left undetermined.
They both originate from the brane early Universe but can be in principle independently determined;
the former is a problem of gravitational collapse and the latter is that of relativistic hydrodynamics on the gravitational background.
$ \alpha_\mathrm{i} $ is directly connected with the primordial density perturbation spectra, and can be potentially given under specific braneworld models, such as the bulk inflaton model \cite{BulkInflaton} or the cyclic Universe model \cite{Cyclic} and so on.
Note that our results will not be qualitatively changed even if the assumption of constant $ \alpha_\mathrm{i} $ is moderately violated;
in the both analyses, what was important was only an absolute value of the mass spectrum for narrow mass range.

In the analyses, we assumed that an evaporating black hole holds the same temperature as the initial value through its lifetime.
Hence the effective total degree of freedom, $ g_\mathrm{eff} $, was fixed at a constant, which is for zero temperature.
However, as was shown by MacGibbon \cite{MacGibbon1991}, the degree increases maximally by an order of magnitude.
Due to such an underestimate of evaporation rate of hot PBHs, our resultant antiproton flux was somewhat overestimated and constraints were in turn virtually strengthened.

We should make a remark on the range of viability of our methodology presented in this and the previous paper \cite{SendoudaNagatakiSato2003} which served constraints on a higher-dimensional cosmology.
Although our interest has been throughout concentrated on a particular cosmological model of RS2 braneworld, we insist that the use of Hawking radiation from small black holes is available for a broad range of models with large extra dimensions.
The most crucial feature for the purpose is the change of Hawking-radiation spectra emerging when a hole notices extra space extending around.
This will generally occurs if only the target theory has compact extra dimensions and the hole shrinks enough below the compactification scale.
This compactification is not necessarily realized by periodic boundary conditions, but RS2-like effective localization of gravity is also permissible.
Examples of phenomenological models contain Arkani-Hamed-Dimopoulos-Dvali \cite{ADD} and Ho\v{r}ava-Witten cosmologies \cite{HW}.
Further, if the theory has a fundamental mass scale lower than 4D Planck, which is somewhat likely, then Hawking temperature will drops like that in the RS2 analysis and hence our results may be directly applied with some modification.

Here we summarize this paper.
We investigated antiproton flux emitted from PBHs distributed over the Galaxy as dark matter.
Mass of PBHs which could contribute to the antiproton spectrum was in the 5D evaporating region of the present mass spectrum, which meant that the effect of accretion had been almost completely swept out.
Because the antiproton emissivity of individual PBH was almost unchanged from that in the 4D case, the change in the resultant antiproton flux only came from the change of the slope of the mass spectrum;
in the large extra-dimension case, $ \mathrm{d}n/\mathrm{d}M \sim M $, while with the small extra-dimension, $ \mathrm{d}n/\mathrm{d}M \sim M^2 $.
However, this also does not affect the antiproton flux so much because the flux is still dominated by the contribution of PBHs with temperatures around $ 1~\text{GeV} $.
Finally we understood that the whole situation is almost identical with the 4D case, which was shown by the expected flux in Fig.~\ref{fig:pbar.total}.
Therefore we expect that even future high-precision experiments are done in sub-GeV kinetic-energy region, it will be difficult to identify braneworld signatures from the spectral shape of antiproton flux only;
this is the crucial difference from the photon case obtained in our previous paper \cite{SendoudaNagatakiSato2003}.

As the most impressive result, we found that the antiproton flux from brane PBHs was a decreasing function of the bulk curvature radius $ l $ like
\begin{equation}
\Phi_\pbar \propto \alpha_\mathrm{i} l^{-13/16+3\F/2}.
\end{equation}
The reason why the flux decreases is that the temperature of a black hole is lowered as $ l $ enlarges, while accretion takes a quite limited role in contrast to the photon case \cite{SendoudaNagatakiSato2003}.
Therefore lower bounds on the size of the extra dimension were obtained.
Of course, on the contrary, constraints on $ \alpha_\mathrm{i} $ from antiprotons could be relaxed by the factor of $ l^{13/16-3\F/2} $.
After all, allowed regions of the braneworld parameters from antiproton observation were presented in Figs.~\ref{fig:pbar.region.0}--\ref{fig:pbar.region.2} for each confidence level.

\begin{figure}[ht]
\begin{center}
\scalebox{0.7}{
\includegraphics[width=12cm,clip]{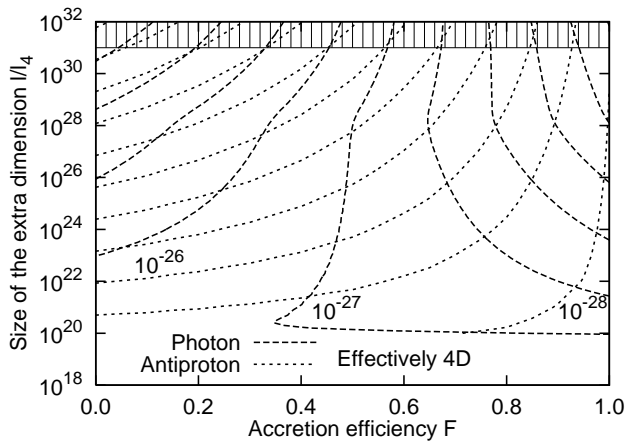}
}
\end{center}
\caption{\label{fig:both.region}
The composite of the two results of allowed regions obtained in Sec.~\ref{sec:pbar} and in our previous paper for high-energy diffuse photons \cite{SendoudaNagatakiSato2003} (presentation is a bit refined).
Regions allowed by both observations is in the left-hand side of each boundary.
The density enhancement in the Galaxy $ \G $ is set to $ 10^5 $ here.
Numbers in the figure indicate $ \alpha_\mathrm{i} $'s.
For $ \alpha_\mathrm{i} \gtrsim 10^{-26} $, diffuse photon constraints are stronger than that of antiprotons.
However, for $ \alpha_\mathrm{i} \sim 10^{-26}\text{--}10^{-28} $, antiprotons set lower bounds on the size of the extra dimension, which was not done by diffuse photons.
For $ \alpha_\mathrm{i} \lesssim 10^{-28} $, antiprotons do almost nothing.
}
\end{figure}

Finally we show the composite of the allowed regions obtained from two distinctive observations, i.e., above-mentioned antiprotons and the diffuse photon background \cite{SendoudaNagatakiSato2003};
see Fig.~\ref{fig:both.region}.
In Sec.~\ref{sec:pbar}, the density enhancement of the Galaxy $ \G $ was left undetermined, but we here use a conservative value $ \G \sim 10^5 $.
Thus the boundaries for antiprotons in Fig.~\ref{fig:both.region} now correspond to $ \alpha_\mathrm{i} = 10^{-18} \text{--} 10^{-28} $.
We find two facts from the figure:
(i) for boundaries corresponding to $ \alpha_\mathrm{i} \gtrsim 10^{-26} $, constraints from diffuse photons are always more strict than that from antiprotons, while (ii) for boundaries of $ \alpha_\mathrm{i} \sim 10^{-26} \text{--} 10^{-28} $, they have chance to cooperate and set the lower bound of $ l $.

In the given situation, the strongest implication is obtained when we assume that the apparent low-energy enhancement of the antiproton flux is actually due to the PBH evaporation.
If so, it means that the PBH abundance $ \alpha_\mathrm{i} $ is near the upper bound, at most one order of magnitude below it.
On the other hand, we here should remind that there is another requirement that diffuse photons from PBHs must be hidden in the background component.
Referring to Fig.~\ref{fig:both.region}, such a condition is interpreted as follows.
Sub-critical PBH abundance ($ \alpha_\mathrm{i} \sim \mathrm{LIM_i} $) means that allowed braneworld parameters are nearly restricted in a belt-shaped region expanding slightly above antiproton-lower bounds, i.e., dotted lines.
On the other hand, the parameters must be in a corresponding allowed region determined by photons, which is on the left-hand side of dashed lines.
We find that such two conditions are only satisfied when $ \alpha_\mathrm{i} \approx 10^{-26}\text{--}10^{-28} $.
Furthermore, it means $ l \lesssim 10^{24} $.
This upper bound is about seven orders of magnitude stronger than that obtained in today's experiments, and may become a disadvantage of the braneworld scenarios.
In the future, more contamination-free observations of PBHs, such as antideuterons or other anti nuclei, may give an answer to the problem.

\begin{acknowledgments}
The authors are all pleased to acknowledge helpful comments by Akira Yamamoto.
The authors would also like to thank Hiroaki Aihara, Tomoyuki Sanuki and Hideyuki Fuke for useful comments.
This work was supported in part through Grant-in-Aid for Scientific Research (S) No.~14102004, Grant-in-Aid for Scientific Research on Priority Areas No.~14079202 and Grant-in-Aid for Scientific Research No.~16740134 by MEXT (Ministry of Education, Culture, Sports, Science and Technology).
Y.S. and K.K.'s work is partly supported by JSPS (Japan Society for the Promotion of Science).
\end{acknowledgments}

\begin{appendix}

\section{\label{sec:pbh}
Shapes of the PBH mass spectrum
}

\subsection{Initial spectrum}

To begin with, we calculate the {\em initial} mass spectrum.
The first case is with mass $ M_\mathrm{i} \lesssim f M_\mathrm{c} $.
PBHs with such light mass are formed in the brane high-energy phase and experience the accretion.
From Eqs.~(\ref{eq:brane.cosmology}), (\ref{eq:pbh.initialmass}), and (\ref{eq:pbh.initialspectrum.def}), the comoving mass spectrum of PBHs with initial mass $ M_\mathrm{i} $ is obtained as
\begin{align}
\frac{\mathrm{d}n}{\mathrm{d}M_\mathrm{i}}
 & = \alpha_\mathrm{i}(M_\mathrm{i})
     \frac{\rho_\mathrm{rad}(t_\mathrm{i})}{M_\mathrm{i}}
     \frac{a(t_\mathrm{i})^3}{M_\mathrm{i}} \nonumber \\
 & = \frac{3 M_4^3}{2^{17/4} \pi}
     \frac{a_\mathrm{eq}^3}
          {t_\mathrm{eq}^{3/2}}
     \alpha_\mathrm{i}(M_\mathrm{i}) f^{1/8} l^{-3/8} M_\mathrm{i}^{-17/8}
     \nonumber \\
 & \qquad \qquad \text{for} \quad M_\mathrm{i} \lesssim f M_\mathrm{c}.
\label{eq:pbh.initialspectrum}
\end{align}
This has a slightly red spectral index in comparison with the 4D case.
For four-dimensional PBHs with $ M_\mathrm{i} \gtrsim f M_\mathrm{c} $, from Eqs.~(\ref{eq:brane.cosmology}), (\ref{eq:pbh.initialmass}), and (\ref{eq:pbh.initialspectrum.def}),
\begin{align}
\frac{\mathrm{d}n}{\mathrm{d}M_\mathrm{i}}
 & = \frac{3 M_4^3}{32\pi}
     \frac{a_\mathrm{eq}^3}
          {t_\mathrm{eq}^{3/2}}
     \alpha_\mathrm{i}(M_\mathrm{i}) f^{1/2} M_\mathrm{i}^{-5/2} \nonumber \\
 & \qquad \qquad \text{for} \quad M_\mathrm{i} \gtrsim f M_\mathrm{c}.
\end{align}
This mass dependence is actually the same as that of the pure 4D case.

\subsection{
Primordial spectrum
}

The next task we have to do is to consider the effects of accretion \cite{GCL2002b,Majumdar2003,SendoudaNagatakiSato2003}.
Here we call the quantities which take accretion into consideration but does not evaporation as ``primordial'' ones.

5D black holes are born before $ t_\mathrm{c} $ and their masses will increase due to accretion during the brane high-energy era.
Resulting primordial mass $ M_\mathrm{p} $ after accretion is obtained from Eqs.~(\ref{eq:brane.cosmology}) and (\ref{eq:pbh.initialmass}) as
\begin{equation}
M_\mathrm{p}
 = \left(\frac{t_\mathrm{c}}{t_\mathrm{i}}\right)^{2 F/\pi} M_\mathrm{i}
 = \left(4 M_4^2 \frac{f l}{M_\mathrm{i}}\right)^{F/\pi} M_\mathrm{i}.
\end{equation}
Combining it with Eq.~(\ref{eq:pbh.initialspectrum}), we obtain the primordial PBH mass spectrum as
\begin{align}
\frac{\mathrm{d}n}{\mathrm{d}M_\mathrm{p}}
 & = \frac{\mathrm{d}n}{\mathrm{d}M_\mathrm{i}}
     \frac{\mathrm{d}M_\mathrm{i}}{\mathrm{d}M_\mathrm{p}} \nonumber \\
 & = \frac{3 M_4^{9/4}}{2^{17/4}\pi}
     \frac{a_\mathrm{eq}^3}
          {t_\mathrm{eq}^{3/2}}
     \alpha_\mathrm{i}[M_\mathrm{i}(M_\mathrm{p})]
     \left(1+\frac{8\F}{9}\right) (4 M_4^2)^\F \nonumber \\
 & \quad \times f^{1/8+\F} l^{-3/8+\F} M_\mathrm{p}^{-17/8-\F} \nonumber \\
 & \qquad \qquad \text{for} \quad M_\mathrm{p} \lesssim f M_\mathrm{c},
\label{eq:pbh.primordialspectrum}
\end{align}
where a new accretion parameter $ \F \equiv 9F/(8\pi-8F) $ was introduced for convenience.
Particularly we find a relation between the distorted spectrum and that with no accretion as follows:
\begin{equation}
\frac{\mathrm{d}n}{\mathrm{d}M_\mathrm{p}}
 = \frac{\alpha_\mathrm{i}(M_\mathrm{i})}{\alpha_\mathrm{i}(M_\mathrm{p})}
   \left(4 M_4^2 \frac{f l}{M_\mathrm{p}}\right)^{\F}
   \left.\frac{\mathrm{d}n}{\mathrm{d}M_\mathrm{p}}\right|_{\text{no accretion}}
\end{equation}

In the $ M_\mathrm{i} \gtrsim f M_\mathrm{c} $ case, they are born in the standard radiation-dominated era and do not experience accretion, so this part of the primordial mass spectrum is unchanged from the initial spectrum.
Hence, identifying $ M_\mathrm{p} \equiv M_\mathrm{i} $,
\begin{equation}
\frac{\mathrm{d}n}{\mathrm{d}M_\mathrm{p}}
 = \frac{\mathrm{d}n}{\mathrm{d}M_\mathrm{i}}
   \quad \text{for} \quad M_\mathrm{p} \gtrsim f M_\mathrm{c},
\end{equation}
where $ \mathrm{d}n/\mathrm{d}M_\mathrm{i} $ is given in Eq.~(\ref{eq:pbh.initialspectrum}).

\subsection{\label{sec:pbh.presentspectrum}
Present spectrum
}

The present mass spectrum is a bit complicated because evaporation is taken into account.
After accretion on PBHs ends, PBHs begin to radiate and lose their mass.
An important feature is that the mass decreasing rate of each PBH is different according to its dimensionality.

In order to obtain the complete mass spectrum, we argue detailed evolution of each PBH.
We can understand that at the present age of the Universe, there are four destinies prepared for each PBH according to its primordial mass $ M_\mathrm{p} $:
(i) born as a 4D PBH and now still be 4D one (their mass is symbolically denoted as $ M_\mathrm{p}^\mathrm{i} $),
(ii) born as a 4D PBH but now is 5D one ($ M_\mathrm{p}^\mathrm{ii} $),
(iii) born as a 5D PBH and now evaporating ($ M_\mathrm{p}^\mathrm{iii} $),
and
(iv) born as a 5D PBH and having already evaporated away ($ M_\mathrm{p}^\mathrm{iv} $).
The mass hierarchy among above typical values is
\begin{equation}
M_\mathrm{p}^\mathrm{i}
 > M_\mathrm{p}^\mathrm{ii}
 > M_\mathrm{p}^\mathrm{iii}
 > M_\mathrm{p}^\mathrm{iv}.
\end{equation}
What is non-trivial is the relations between every typical mass and $ M_\mathrm{p}^* $, the primordial mass of a PBH with lifetime equal to the age of the Universe $ t_0 $.
If a condition $ M_\mathrm{p}^* > M_\mathrm{p}^\mathrm{iii} $ holds, PBHs which were originally five-dimensional have already died out and there are only originally four-dimensional ones now.
To the contrary, in the case of $ M_\mathrm{p}^* < M_\mathrm{p}^\mathrm{iii} $, there are still originally 5D PBHs, which have experienced accretion in the brane early Universe.

Here we note that our framework, braneworld, itself has a mass scale $ M_\mathrm{c} $, which determines whether a BH is five dimensional or four dimensional according to $ M \lessgtr M_\mathrm{c} $.
If we require that PBHs having just evaporated were originally five-dimensional, we need $ M_\mathrm{c} > M_\mathrm{p}^* $ and hence
\begin{equation}
l \gtrsim \left(g_\mathrm{eff,5} t_0\right)^{1/3} \sim 10^{20}.
\label{eq:pbh.moderate}
\end{equation}
This condition is in general important for constraints from cosmic-rays since, if it does not hold, we cannot see any braneworld signature in those observations;
see Sec.~\ref{sec:pbar} (there a bit stringent condition was required) and our previous paper \cite{SendoudaNagatakiSato2003}.
Additionally, there is another mass scale implemented in the framework, $ M_\mathrm{c}^* $, which is defined as
\begin{equation}
M_\mathrm{c}^* \equiv \sqrt{M_\mathrm{c}^2-\frac{g_\mathrm{eff,5}t_0}{l}}.
\end{equation}
Recalling Eq.~(\ref{eq:pbh.evap}), it is understood that a PBH presently with $ M_\mathrm{c}^* $ has primordial mass $ M_\mathrm{p} = M_\mathrm{c} $.
If the mass of a PBH is now $ M < M_\mathrm{c}^* $, we can see that it was born as 5D one because the condition means that its primordial mass $ M_\mathrm{p} $ was smaller than $ M_\mathrm{c} $.

Now let us investigate the resultant mass spectrum.
We already understood that PBHs in the mass range smaller than $ M_\mathrm{p}^* $ are all on the way to evaporation.
Combining the expression of the primordial mass spectrum Eq.~(\ref{eq:pbh.primordialspectrum}) and the mass evolution formula Eq.~(\ref{eq:pbh.evap}), we can calculate the present mass spectrum, which is formally written as
\begin{equation}
\frac{\mathrm{d}n}{\mathrm{d}M}
 = \frac{\mathrm{d}n}{\mathrm{d}M_\mathrm{i}}
   \frac{\mathrm{d}M_\mathrm{i}}{\mathrm{d}M_\mathrm{p}}
   \frac{\mathrm{d}M_\mathrm{p}}{\mathrm{d}M}.
\label{eq:pbh.formalspectrum}
\end{equation}
For the three types of PBH (i)-(iii), we can calculate derivative factors in the above equation and finally obtain whole mass spectrum as follows.
Note that we fix $ f = 1 $ for simplicity.

\subsubsection{$ M \gtrsim M_\mathrm{c} $}

The present mass of the black hole is larger than boundary value.
It is now four-dimensional and was born as 4D one.
Its history is described as
\begin{equation}
M_\mathrm{p} = M_\mathrm{i}
\quad \text{and} \quad
M(t) = \left[M_\mathrm{p}^3 - g_\mathrm{eff,4}(t-t_\mathrm{i})\right]^{1/3}.
\end{equation}
Thus the present mass spectrum of this region becomes
\begin{align}
\frac{\mathrm{d}n}{\mathrm{d}M}
 & = \frac{3}{32 \pi}
     \frac{a_\mathrm{eq}^3}
          {t_\mathrm{eq}^{3/2}}
     \alpha_\mathrm{i}[M_\mathrm{i}(M)]
     f^{1/2} M^2 \left[M^3 + g_\mathrm{eff,4} t_0\right]^{-3/2}
     \nonumber \\
 & \qquad \qquad \text{for} \quad M \gtrsim M_\mathrm{c}.
\end{align}

\subsubsection{$ M_\mathrm{c} \gtrsim M \gtrsim M_\mathrm{c}^* $}

Next is the most complicated case;
large initial mass, but not enough.
A PBH in this category was born as a 4D one and now five-dimensional.
Such ones only exist if $ l \sim 10^{19} $, so usually we do not have to pay attention to them.
The history is
\begin{equation}
M_\mathrm{p} = M_\mathrm{i}
\end{equation}
and
\begin{equation}
M(t)
 = \left\{
 \frac{g_\mathrm{eff,5}}{l}
 \left[\frac{1}{g_\mathrm{eff,4}}
       \left(M_\mathrm{p}^3-M_\mathrm{c}^3\right)
        - t\right]
 + M_\mathrm{c}^2
   \right\}^{1/2}.
\end{equation}
The present mass spectrum is
\begin{align}
\frac{\mathrm{d}n}{\mathrm{d}M}
 & = \frac{1}{16 \pi}
     \frac{a_\mathrm{eq}^3}
          {t_\mathrm{eq}^{3/2}}
     \alpha_\mathrm{i}[M_\mathrm{i}(M)] f^{1/2}
     \frac{g_\mathrm{eff,4}}{g_\mathrm{eff,5}} l \nonumber \\
 & \quad \times M \left[\frac{g_\mathrm{eff,4}}{g_\mathrm{eff,5}} l
             \left(M^2-M_\mathrm{c}^2\right)
             + M_\mathrm{c}^3 + g_\mathrm{eff,4} t_0\right]^{-3/2}
   \nonumber \\
 & \qquad \qquad \text{for} \quad M_\mathrm{c} \gtrsim M \gtrsim M_\mathrm{c}^*.
\label{eq:pbh.presentspectrum2}
\end{align}

\subsubsection{$ M_\mathrm{c}^* \gtrsim M $}

A PBH of this type is sufficiently light and was natively five-dimensional.
Readers should remind that such a kind of PBH can only exist under a particular condition on the bulk radius, Eq.~(\ref{eq:pbh.moderate}).
The history is
\begin{equation}
M_\mathrm{p} = (4fl)^{F/\pi} M_\mathrm{i}^{(\pi-F)/\pi}
\end{equation}
and
\begin{equation}
M(t)
 = \left[M_\mathrm{p}^2 
         - \frac{g_\mathrm{eff,5}}{l}(t-t_\mathrm{c})\right]^{1/2}.
\end{equation}
Hence the present mass spectrum is
\begin{align}
\frac{\mathrm{d}n}{\mathrm{d}M}
 & = \frac{3}{2^{17/4}\pi}
     \frac{a_\mathrm{eq}^3}
          {t_\mathrm{eq}^{3/2}}
     \alpha_\mathrm{i}[M_\mathrm{i}(M_\mathrm{p})]
     \left(1+\frac{8\F}{9}\right) \nonumber \\
 & \quad \times 4^\F f^{1/8+\F} l^{-3/8+\F} M
   \left[M^2+\frac{g_\mathrm{eff,5}}{l}t_0\right]^{-25/16-\F/2} \nonumber \\
 & \qquad \qquad \text{for} \quad M_\mathrm{c}^* \gtrsim M.
\label{eq:pbh.presentspectrum3}
\end{align}

\section{\label{sec:pbar.propagation}
Propagation of antiprotons in the Galaxy
}

In this paper, we adopt a semi-analytical method which was first proposed by Webber, Lee, and Gupta \cite{Webberetal1992} and has been extensively utilized by other authors \cite{Barrauetal2002,Donatoetal2001,Maurinetal2001,Maurinetal2002,TailletMaurin2003,MaurinTaillet2003}.
This method is based on the diffusion model proposed by Ginzburg, Khazan, and Ptuskin \cite{Ginzburgetal1980}.

At the beginning, we write down the full diffusion equation in the steady-state for the differential number density $ N(r,z,E) \equiv \mathrm{d}n_\pbar/\mathrm{d}E $ \cite{Berezinskiietal1990}:
\begin{multline}
0 = \frac{\partial N}{\partial t}
  = \vec{\nabla}
    \cdot [K(E) \vec{\nabla} N(r,z,E) - \vec{V}_\mathrm{c}(r,z) N(r,z,E)] \\
  + \frac{\vec{\nabla}\cdot\vec{V}_\mathrm{c}(r,z)}{3}
    \frac{\partial}{\partial E}
    \left[\frac{p^2}{E}N(r,z,E)\right]
  - \Gamma(E) N(r,z,E) \\
  + \Q(r,z,E)
  + \frac{\partial}{\partial E}
    \left\{-\left[b_\mathrm{reacc}(E)+b_\mathrm{coll}(E)\right] N(r,z,E)
    \right\} \\
  + \frac{\partial}{\partial E}
    \left[\beta^2 K_{pp}(E) \frac{\partial}{\partial E} N(r,z,E)\right].
\label{eq:pbar.diffusion}
\end{multline}
As was mentioned in Sec.~\ref{sec:pbar}, we assume cylindrical symmetry for the Galaxy, and expand all the physical quantities in terms of the Bessel functions of order zero.
Such quantities, for example, differential number density $ N(r,z,E) $ or the source term $ \Q(r,z,E) $, can be expanded as
\begin{equation}
f(r,z,E) = \sum_{i=1}^\infty f_i(z,E) J_0\left(\zeta_i \frac{r}{R}\right),
\end{equation}
where the components are given by
\begin{equation}
f_i(z,E)
 = \frac{2}{[J_1(\zeta_i)]^2}
   \int\limits_0^R \frac{r}{R} f(r,z,E)
   J_0\left(\zeta_i \frac{r}{R}\right) \frac{\mathrm{d}r}{R},
\end{equation}
and $ \zeta_i $'s are the zeros of $ J_0 $.
As a consequence, the problem has been reduced to diffusion along the vertical direction $ z $.
From now, let us see the roles of each term in the diffusion equation Eq.~(\ref{eq:pbar.diffusion}).

\subsection{Diffusion and convection}

$ K(E) $ and $ \vec{V}_\mathrm{c} $, which appeared in the first and second term in the diffusion equation, stand for diffusion and convection, respectively.
During the propagation in the halo, charged particles are diffused by the random galactic magnetic field.
We adopt standard discussions  \cite{Maurinetal2001,Donatoetal2001,Maurinetal2002,BarrowMaartens2002,TailletMaurin2003,MaurinTaillet2003,Ptuskinetal1997} for it;
the energy-dependent diffusion coefficient is given as
\begin{equation}
K(E) = K_0 \beta \R^\delta,
\end{equation}
where $ \beta $ denotes the velocity and $ \R \equiv p/Z $ is the rigidity.
For antiprotons, $ \R $ is identical with its momentum $ p $.
Here, in principle $ K_0 $ is only an undetermined parameter since $ \delta $ should be fixed by microphysics.
We have set $ \delta $ to 0.6 throughout this paper in accordance with previous results \cite{Maurinetal2001,Donatoetal2001,Maurinetal2002,BarrowMaartens2002,TailletMaurin2003,MaurinTaillet2003}.

Another potentially important effect for cosmic-ray transfer is convection.
The convective (or galactic) wind, which consists of plasma medium, flows outward from the galactic disc \cite{Parker1965}.
For simplicity, we assume that it holds a constant value in the whole region of the diffusion halo, i.e., the convective wind $ \vec{V}_\mathrm{c} $ takes the form
\begin{equation}
\vec{V}_\mathrm{c}(r,z) = V_\mathrm{c} \frac{\vec{z}}{|z|},
\end{equation}
where $ V_\mathrm{c} $ is a parameter to be fixed.
The convective flow also contributes to the energy loss in the third term through the adiabatic expansion of the plasma.
This energy loss takes place only in the region where its divergence has a non-zero value, namely on the galactic disc.
The adiabatic energy-loss expressed by the third term is eventually written down as
\begin{equation}
b_\mathrm{adiab}
 = -2 h \delta(z) T \left(\frac{T+2m}{T+m}\right) \frac{V_\mathrm{c}}{3h},
\end{equation}
where $ T $ and $ m $ are kinetic energy and rest mass of antiprotons, respectively.

\subsection{\label{sec:pbar.propagation.interaction}
Inelastic interactions
}

The fourth term represents the total rate of interactions with the interstellar gas by which an antiproton escapes from the energy bin.
Such interactions occur only in the galactic disc.
For collisions between antiprotons and hydrogen nuclei, $ \Gamma(E) $ is defined as
\begin{equation}
\Gamma(E) = 2 h \delta(z) \Gamma_{\pbar p}(E),
\end{equation}
where
\begin{equation}
\Gamma_{\pbar p}(E)
 = (n_\mathrm{H} + 4^{2/3} n_\mathrm{He}) \beta \sigma_{\pbar p}(E).
\end{equation}
$ \sigma_{\pbar p} $ is the total cross section of the interaction of interest.

\subsection{Energy gain -- reacceleration}

During propagation, antiprotons are reaccelerated by the random magnetic field.
The effect of the Fermi acceleration appears through the diffusion coefficient in the energy space defined as \cite{Maurinetal2002}
\begin{equation}
K_{pp}(E)
 = \frac{4}{3 \delta (4-\delta^2) (4-\delta)}
   \frac{V_\mathrm{A}^2 p^2}{K(E)},
\end{equation}
where $ V_\mathrm{A} $ is the \Alfven velocity of the galactic magnetic field.

Two terms in the diffusion equation contain $ K_{pp} $:
one is the coefficient of the first derivative of energy
\begin{equation}
b_\mathrm{reacc}
 = 2 h \delta(z) \frac{1+\beta^2}{E} K_{pp}(E),
\end{equation}
and another is in the second derivative of energy.

\subsection{Energy losses}

We take into account three types of energy-loss for $ b_\mathrm{coll} $: two collisional processes, namely Coulomb scattering and ionization, and one adiabatic process.
For these energy losses, see \cite{MannheimSchlickeiser1994,Nordgrenetal1992}.

\subsection{Sources}

Three kinds of the sources have been described in Sec.~\ref{sec:pbar}.
With the thin-disc approximation on the Galaxy, the source term takes a form as
\begin{align}
\Q_i(z,E)
 & = \Q_i^\mathrm{pri}+\Q_i^\mathrm{sec}+\Q_i^\mathrm{ter}[N_i] \nonumber \\
 & = q_i^\mathrm{pri}
     + 2 h \delta(z) q_i^\mathrm{sec}
     + 2 h \delta(z) q_i^\mathrm{ter}[N_i],
\end{align}
where
\begin{align}
q_i^\mathrm{pri}(z,E)
 & = \int \frac{\mathrm{d}^2\tilde{N}_\pbar}{\mathrm{d}E\mathrm{d}t}(E,M)
     \frac{\mathrm{d}\tilde{n}_i}{\mathrm{d}M}(z,M) \mathrm{d}M,
\label{eq:pbar.pri.Bess} \\
q_i^\mathrm{sec}(0,E)
 & = \int\limits_{E_\mathrm{th}}^\infty \mathrm{d}E_p
     \beta N_{p,i}(E_p)
     \sum_{A=\mathrm{H,He}} n_A
     \frac{\mathrm{d}\sigma_{pA \rightarrow \pbar X}}
          {\mathrm{d}E}(E_p, E),
\label{eq:pbar.sec.Bess}
\end{align}
and
\begin{multline}
q_i^\mathrm{ter}[N_i](0,E) \\
 = (n_\mathrm{H} + 4^{2/3} n_\mathrm{He})
   \int\limits_{E}^\infty \mathrm{d}E'
   \frac{\sigma_{\pbar p}^\mathrm{non-ann}(T')}{T'}
   \beta N_i(z,E') \\
 - (n_\mathrm{H} + 4^{2/3} n_\mathrm{He})
   \sigma_{\pbar p}^\mathrm{non-ann}(T) \beta N_i(z,E).
\label{eq:pbar.ter.Bess}
\end{multline}
$ \mathrm{d}\tilde{n}_i/\mathrm{d}M $'s in Eq.~(\ref{eq:pbar.pri.Bess}) are Bessel components of $ \mathrm{d}\tilde{n}/\mathrm{d}M $.
A relation between flux and number density
\begin{equation}
\Phi(r,z,E) = \frac{\beta}{4\pi} N(r,z,E)
\end{equation}
was made use of.

\subsection{Final form of the diffusion equation}

After above considerations, the final form of the diffusion equation for the antiproton number density becomes
\begin{multline}
0 = \left[
 K(E) \frac{\partial^2}{\partial z^2}
 - V_\mathrm{c} \frac{\partial}{\partial z}\right] N_i(z,E) \\
 - \left[K(E) \frac{\zeta_i^2}{R^2}
          + 2 h \delta(z) \Gamma_{\pbar p}(E)\right] N_i(z,E) \\
 + \Q_i(z,E)
 + \left\{\frac{\partial}{\partial E}
         \left[-b(E) + \beta^2 K_{pp}(E)
                        \frac{\partial}{\partial E}\right]
   \right\} N_i(z, E),
\end{multline}
where $ b = b_\mathrm{reacc} + b_\mathrm{coll} + b_\mathrm{adiab} $.
This is the basic equation to be solved.

\end{appendix}


\end{document}